\documentclass{aastex63}
\shorttitle{C and O isotopic ratios for variable M III}
\shortauthors{Lebzelter et al.}

\begin{document}

\title{CARBON AND OXYGEN ISOTOPIC RATIOS.  II. SEMI-REGULAR VARIABLE M GIANTS}

\author[0000-0002-0702-7551]{THOMAS LEBZELTER}
\affil{Department of Astrophysics, University of Vienna\\ 
T{\"u}rkenschanzstrasse 17, 1180 Vienna, Austria}
\email{thomas.lebzelter@univie.ac.at}

\author[0000-0002-2726-4247]{KENNETH H. HINKLE}
\affil{NSF's National Optical-Infrared Astronomy Research Laboratory\\
P.O. Box 26732, Tucson, AZ 85726, USA}
\email{khinkle@noao.edu}

\author[0000-0002-5514-6125]{OSCAR STRANIERO}
\affil{INAF, Osservatorio Astronomico d'Abruzzo.\\ 
I-64100 Teramo, Italy, and INFN-LNGS, Assergi (AQ), Italy}
\email{straniero@oa-teramo.inaf.it}

\author[0000-0003-1814-3379]{DAVID L. LAMBERT}
\affil{W.J. McDonald Observatory and Department of Astronomy\\
The University of Texas at Austin, Austin, TX 78712-1205, USA}

\author[0000-0002-3007-206X]{CATHERINE A. PILACHOWSKI}
\affil{Astronomy Department, Indiana University Bloomington\\
Swain West 319, 727 East Third Street, Bloomington, IN 47405-7105, USA}

\author[0000-0001-8383-5115]{KRISTIE A. NAULT}
\affil{Department of Physics \& Astrononomy, Univeristy of Iowa\\
203 Van Allen Hall, Iowa City, IA 52242-1479, USA}

\begin{abstract}
Carbon and oxygen isotopic ratios are reported for a sample of 51 SRb- and Lb-type variable asymptotic giant branch stars.  
Vibration-rotation first- and second-overtone CO lines in 1.5 to 2.5 $\mu$m spectra were 
measured to derive isotopic ratios for $^{12}$C/$^{13}$C, $^{16}$O/$^{17}$O, and $^{16}$O/$^{18}$O.  
Comparisons with  previous measurements for individual stars and with various samples of evolved stars, 
as available in the extant literature, are discussed.  Using the oxygen isotopic ratios, the masses of 
the SRb stars can be derived.  Combining the masses with Gaia luminosities, the SRb stars are shown 
to be antecedents of the Mira variables.  The limiting parameters where plane-parallel, hydrostatic equilibrium 
model atmospheres can be used for abundance analysis of M giants are explored. 
\end{abstract}


\section{INTRODUCTION}
The upper giant branch is a decisive phase in the stellar evolution of low- and intermediate-mass stars. 
Large radial extension, low effective temperature, and stellar pulsation lead to significant mass loss. 
The material lost to the interstellar medium (ISM) in this way has been enriched by products of the stellar nucleosynthesis resulting from deep convective mixing (dredge-up). 
Therefore, these objects play an important role in the cosmic matter cycle.  Abundance analysis of the surface composition probes
the conditions and processes in the stellar interior.  

The dredge-up replaces the primordial composition of the C, N, and O with the post-hydrogen-burning one. For the sun, the surface values are those of the presumed typical primordial gas, $^{12}$C/$^{13}$C = 89 \citep{meibom_etal_2007}, $^{16}$O/$^{17}$O = 2700, and $^{16}$O/$^{18}$O = 498 \citep{lodders_etal_2009}.
The first major alteration of the surface composition occurs during the first dredge-up at the bottom of the red giant branch.
Depending on the stellar parameters, further mixing events take place, in particular with the onset of double-shell burning on the asymptotic giant branch (AGB), the so-called third dredge-up.

This is our second Paper In a series to study the isotopic ratios of C and O in AGB stars. 
In the first paper of this series \citep[][Paper I]{hinkle_et_al_2016} we presented an extensive discussion on the influence of stellar parameters and evolution on the isotopic ratios of these two key elements measured at the stellar surface.
Here we confine the discussion to a brief summary of the main effects and refer to Paper I (note in particular Figure 10), \citet{lebzelter_et_al_2015}, and references therein for a more extensive description.

The CNO cycle 
depletes $^{12}$C and enhances $^{13}$C, a change that becomes visible as a steep drop in 
the $^{12}$C/$^{13}$C ratio down to 10 to 25 after the first dredge up.
A slight dependence of the resulting ratios on stellar mass is expected, with the more massive stars showing higher $^{12}$C/$^{13}$C values (Paper I).
Extensive observational evidence \citep[e.g.][]{charbonnel_et_al_1999} exists that extra-mixing along the red giant branch (RGB) can further reduce the ratio between the two carbon isotopes by a factor of two in low-mass stars.
The interplay between production, mixing, and destruction of the various oxygen isotopes results in a dependence of $^{16}$O/$^{17}$O on stellar mass \citep{boothroyd_et_al_1994, eleid_1994, karakas_lattanzio_2014}, with the istopic ratio decreasing with increasing mass as different depths  and thus different regions of $^{17}$O enhancement are reached by the first dredge-up.
$^{18}$O, on the other hand, is depleted by H burning and therefore $^{16}$O/$^{18}$O is increasing with the first dredge-up. 
This effect, however, is almost independent of mass.

The second dredge-up has only a mild effect on the isotopic ratios on the surface. 
With the third dredge-up episodes, which are predicted to occur in stars with main-sequence masses between approximately 1.3 and 5\,M$_{\sun}$ at solar metallicity, $^{12}$C/$^{13}$C is strongly enhanced, while there is little effect on the oxygen isotopic ratios. 
In the more massive AGB stars, Hot Bottom Burning (HBB) can effect the oxygen isotopes, in particular the abundance of $^{18}$O which is significantly depleted \citep{lattanzio_et_al_1996}.
As pointed out by \citet{lebzelter_et_al_2015}, the $^{16}$O/$^{18}$O is good indicator of the primordial $^{18}$O abundance in a star, so that star-to-star variations of this ratio also includes information on the primordial abundance scatter of this isotope.

Isotopic ratios of abundant elements like C and O thus provide a valuable and easily accessible tool to study the evolution of low- and intermediate-mass stars along the AGB and to get access to the main-sequence mass and the primordial composition of these objects.
The usage of this method has a long tradition in astrophysics going back to the 1970s \citep{dearborn_1977}.
In particular, the $^{12}$C/$^{13}$C ratio was measured for a large number of objects with various methods. 
Oxygen isotopic ratios received less attention until the landmark papers by \citet{harris_lambert_1984}, \citet{harris_et_al_1985}, and \citet{smith_lambert_1990}.
These advances resulted from the advent of high-resolution near-infrared spectroscopy in stellar astrophysics.
Measurements in the radio regime \citep[e.g.][]{ramstedt_olofsson_2014} and in the far-infrared \citep[e.g.][]{justtanont_et_al_2015} extended the possibilities for studies of isotopic ratios in evolved stars in the past years.

Very high precision isotopic ratios can be determined from presolar grains. 
Comparing the results
from such grains with model predictions has allowed us to identify the kind of stars from which they originated \citep{zinner_1998, nittler_2009}.
We know that most O-rich presolar grains were formed in the cool circumstellar envelope  around red giants stars lived before the birth of the solar system. 
However, a comparison of predictions with isotopic ratios derived from measurements in stars is still incomplete.
This Paper Is part of our effort to close this gap and to derive isotopic ratios of carbon and oxygen for evolved  low- and intermediate-mass stars.

In such stars, isotopic ratios can be studied spectroscopically owing to the isotopic shifts between the isotopologues of the abundant molecule CO. 
Vibration-rotation molecular transitions well suited for this task are found in the near-infrared range between 1 and 5\,$\mu$m.
In K, M, S, and C giants the low stellar effective temperature results in spectral lines of sufficient strength to detect the less abundant isotopologues.

Previous studies dealt with small samples of stars or focused on only one or two of the three isotopic ratios mentioned above.
In our series of papers, we aim at determining all three isotopic ratios for a large sample of AGB stars in a homogeneous and self-consistent way.
With this data set, our goal is to systematically compare model predictions with observations, describe the population on the AGB more properly in terms of mass, and effectively contrast the results from presolar grains with the atmospheres of the dust producers, namely, the AGB stars.
Our project is restricted to oxygen-rich stars, i.e. spectral type M and S.
Carbon-rich stars present a unique set of issues resulting largely from the very rich spectrum.  Studies on large samples have previously been published \citep[e.g.][]{lambert_et_al_1986, abia_et_al_2017}.

Most, if not all, AGB stars are variable, forming the group of long period variables (LPVs).
Their pronounced variability, brightness, and red color provide good indicators to identify these objects in the sky.
However, as the final evolutionary stage for stars in a wide initial mass range, the AGB population of the Galactic field is highly inhomogeneous in terms of mass, age, composition, and variability characteristics.
Based on the latter, LPVs are typically divided into types M (miras), SRa and SRb variables, with the first two groups having large-amplitude light variations with pronounced periodicity, while the third group is characterized by smaller light amplitudes and phases of irregularity.
For the first paper of this series (Paper I) we derived carbon and oxygen isotopic ratios for a group of Mira and SRa variables.  
These stars were attractive candidates in the context of the galactic evolution, because they have prodigious mass loss and are certain to be on the AGB, presumably near the tip.
The isotopic ratios of C and O indicated that the majority of the M-star miras and SRa variables have main-sequence masses of $\le$2\,M$_{\sun}$. 

The present Paper Is a study of the C and O isotopes in small-amplitude variables, primarily in the SRb class.
In Section \ref{sect:obs} we describe the sample for this study, the observational methods, and the applied steps of the data reduction.
Sects.\,\ref{sect:parameters} and \ref{sect:abundances} present the results for the SRb variables including a relation between the CO excitation temperature and the effective temperature of a star.
We use this relation and Gaia distances to constrain atmospheric models for targets and compare isotopic ratios of oxygen determined by curve-of-growth analysis and by spectrum synthesis.
The paper closes with a discussion of the findings (Section \ref{sect:discuss}).

\section{Observations and Reductions}\label{sect:obs}
\subsection{Sample}
As was the case for Paper I, the spectroscopic material analyzed here is drawn entirely from the archives of the Kitt Peak National Observatory (KPNO) 4-meter Mayall Telescope Fourier Transform Spectrometer (FTS).  
For this paper the sample is mainly M and S SRb stars. 
The variable type was obtained from the latest General Catalog of Variable Stars (GCVS).  
In addition to SRb variables, the GCVS class of some program stars is Lb.  These two variable 
classes have similar, if not identical, characteristics \citep[see][]{lebzelter_obbrugger_2009}.
In addition, we included two stars classified as SRa (GZ Peg and DY Gem) and two as SRc (Y Lyn and $\delta ^2$ Lyr) that are in the FTS archive.
GZ Peg has a GCVS amplitude of less than 0.3 mag and thus belongs to variability class SRb rather than to SRa.
A light curve of the star could not be found in the literature.
For DY Gem, an ACVS light curve \citep{pojmanski_et_al_2005} exists that shows very obvious signs of irregularity. 
The amplitude of the main period is rather small, but the star may have a long secondary period.
We believe that DY Gem can be considered an SRb.
The two SRcs are, if classified correctly, supergiants, and thus not AGB stars. 
The classification of Y Lyn as an SRc is based on its spectroscopic luminosity class Ib-II \citep{keenan_1942}. 
However, in various papers Y Lyn is discussed in the course of AGB stars.
\citet{guandalini_busso_2008} derived a luminosity in accordance with the values they found for miras. 
For $\delta ^2$ Lyr luminosity class II was confirmed in the extensive study of supergiants by \citet{levesque_et_al_2005}.
Therefore, we assume that the classification as SRc is correct. 
We decided to derive isotopic ratios for these stars within the present paper, but we handle them separately in the discussion.
The complete set of spectra used for this project is listed in Table \ref{t:obslog}.
For all the sample stars spectra were present in the archive that cover both the $H$ and $K$ windows. 

The archival spectra were originally observed for a variety of projects including abundance studies. 
We will refer to these former studies when comparing our findings with values from the literature (Section \ref{sect:literature}).
Unlike the time series in Paper I,  
the current data set consists mainly of individual observations.  
Where multiple spectra of a star existed, we typically analyzed those with signal-to-noise ratio (S/N) $\gtrsim$ 50 and used the one with the highest S/N ratio for the abundance results. 
The other spectra were used as an estimate of the uncertainties.

\startlongtable
\begin{deluxetable*}{llllllrr}
\tabletypesize{\footnotesize}
\tablecaption{Observations \label{t:obslog}}
\tablewidth{0pt}
\tablehead{ \colhead{HD} & \colhead{HR/BD} & \colhead{Var} &  \colhead{Var} & \colhead{Date} & \colhead{Res.}  & \colhead{S/N K} & \colhead{S/N H} \\
\colhead{ } & \colhead{ } & \colhead{Name} &  \colhead{Type\tablenotemark{a}} & \colhead{} &  \colhead{(cm$^{-1}$)} & \colhead{peak} & \colhead{}
}
\startdata
 6860 & HR 337   & $\beta$ And &   var  & 76 Jun 6     & 0.08 & --\tablenotemark{b} &    \\  
 6860 & HR 337   & $\beta$ And &   var  & 79 Jun 14+15 & 0.04 & 128                 &    \\ 
 7351 & HR 363   & DT Psc      &   SR:  & 77 Jun 29    & 0.17 & 83                  &    \\         
 18191 & HR 867  & RZ Ari      &   SRb  & 76 Aug 17    & 0.09 & 89                  &    \\    
 18884 & HR 911  & $\alpha$ Cet &   Lb:  & 76 Sep 28    & 0.08 & 74                  &    \\     
 20720 & HR 1003 & $\tau^{4}$ Eri &   Lb   & 76 Sep 29    & 0.09 & 65                  &    \\    
 22649 & HR 1105 & BD Cam      &  Lb   & 76 May 14    & 0.21 & 107                 &    \\     
 30959 & HR 1556 & o$^{1}$ Ori      &  SRb  & 76 Aug 23    & 0.08 & 73                  &    \\  
 30959 & HR 1556 & o$^{1}$ Ori      &   SRb  & 78 Jan 22    & 0.10 & 73                  &    \\    
 39225 & HR 2028 &             &   var  &  89 Sep 14   & 0.10 & 85                  & 49 \\  
 44478 & HR 2286 & $\mu$ Gem   &   Lb   & 76 Sep 29    & 0.08 & 30                  &    \\  
 44478 & HR 2286 & $\mu$ Gem   &   Lb   & 79 Sep 5+7+9 & 0.04 & 268                 &    \\  
 49368 &         & V613 Mon    &   SRb: &  87 Apr 14   & 0.07 & 61                  & 25 \\   
 55966 & HR 2742 & VZ Cam      &   Lb:  &  90 Apr 12   & 0.07 & 149                 & 65 \\  
 58521 &         & Y Lyn       &   SRc  & 78 Jan 22    & 0.10 & 81                  &    \\  
 64332 &         & NQ Pup      &   Lb   & 87 Apr 13    & 0.07 & 56                  & 29 \\  
 71250 & HR 3319 & BP Cnc      &   SRb  & 78 Oct 15    & 0.09 & 63                  &    \\    
 94705 & HR 4267 & VY Leo      &   Lb:  &  90 Apr 11   & 0.07 & 67                  & 34 \\ 
 96360 &         & HL UMa      &   SRb  &  87 Apr 14   & 0.07 & 64                  & 30 \\ 
100029 & HR 4434 & $\lambda$ Dra &  SR & 90 Apr 12     & 0.07 & 98                  & 53 \\
102212 & HR 4517 & $\nu$ Vir   &   SRb & 82 Apr 8     & 0.07 & 100                 & 86 \\
106198 & HR 4647 & V335 Hya    &  Lb   & 87 Apr 13    & 0.07 & 111                 & 61 \\ 
112300 & HR 4910 & $\delta$ Vir &   SR   & 76 Jun 18    & 0.09 & 55                  &   \\ 
112300 & HR 4910 & $\delta$ Vir &   SR   & 77 Jun 24    & 0.09 & 70                  &   \\ 
114961 &         & SW Vir      &   SRb  & 76 Jun 18    & 0.09 & 64                  &   \\    
119228 & HR 5154 & IQ UMa      &   SRb  &  90 Apr 11   & 0.07 & 82                  & 50 \\  
121130 & HR 5226 & CU Dra      &   Lb:  & 77 Jun 28    & 0.09 & 64                  &    \\ 
123657 & HR 5299 & BY Boo      &  Lb:  & 76 Aug 18    & 0.09 & 32                  &    \\         
126327 &         & RX Boo      &   SRb  & 76 Jan 14    & 0.09 & 64                  &    \\    
132813 & HR 5589 & RR UMi      &   SRb  & 76 Aug 18    & 0.08 & 58                  &    \\     
133216 & HR 5603 & $\sigma$ Lib &   SRb  & 76 Aug 18    & 0.08 & 94                  &    \\          
146051 & HR 6056 & $\delta$ Oph &   var  & 77 Jun 24    & 0.09 & 62                  &    \\  
148783 & HR 6146 & g Her       &   SRb  & 77 Jun 30    & 0.09 & 90                  &    \\      
149683 &         & R UMi       &  SRb  & 83 Sep 13    & 0.07 & 98                  &    \\ 
154143 & HR 6337 &             &   var  &  90 Apr 12   & 0.07 & 149                 & 65 \\
163990 & HR 6702 & OP Her      &   SRb  & 76 Sep 29    & 0.08 & 43                  &    \\    
168574 & HR 6861 & V4028 Sgr   &   SR:  & 76 Sep 29    & 0.09 & 43                  &    \\    
175588 & HR 7139 & $\delta$2 Lyr &   SRc: & 76 Apr 11    & 0.09 & 91                  &    \\  
175865 & HR 7157 & R Lyr       &  SRb  &  90 Apr 11   & 0.07 & 168                 & 93 \\ 
182917 &         & CH Cyg      &   SR+Z And & 79 Feb 8 & 0.08 & 133                 &    \\         
183439 & HR 7405 & $\alpha$ Vul &   var  & 76 May 16    & 0.18 & 89                  &    \\      
184313 &         & V450 Aql    &   SRb  & 84 Feb 21    & 0.07 & 76                  &    \\  
184313 &         & V450 Aql    &   SRb  & 84 Nov 13    & 0.07 & 85                  &    \\     
196610 & HR 7886 & EU Del      &   SRb  &  90 Apr 12   & 0.07 & 81                  & 35 \\
198026 & HR 7951 & EN Aqr      &   Lb   & 76 May 16    & 0.16 & 51                  &   \\    
198164 &         & CY Cyg      &   Lb   & 76 Jun 19    & 0.23 & 52                  &   \\       
200527 & HR 8062 & V1981 Cyg   &   SRb  & 87 Apr 13    & 0.07 & 86                  & 53 \\ 
202369 & HR 8128 & 29 Cap      &   var? & 76 May 12    & 0.18 & 90                  &    \\    
205730 & HR 8262 & W Cyg       &   SRb  &  77 Apr 4    & 0.10 & 54                  &    \\  
205730 & HR 8262 & W Cyg       &   SRb  & 84 Apr 8     & 0.07 & 113                 &    \\   
209872 &         & SV Peg      &   SRb  & 83 May 21    & 0.07 & 81                  &    \\    
209872 &         & SV Peg      &   SRb  & 84 Nov 13    & 0.07 & 107                 &    \\    
216386 & HR 8698 & $\lambda$ Aqr &   Lb   & 76 Jun 6     & 0.08 & 64                  &    \\ 
216672 & HR 8714 &  HR Peg     &   SRb  & 77 Jun 4     & 0.17 & 115                 &    \\ 
216672 & HR 8714 & HR Peg      &   SRb  & 88 Jul 2     & 0.04 & 67                  &    \\
217906 & HR 8775 & $\beta$ Peg &   Lb   & 76 May 12    & 0.09 & 105                 &    \\    
218634 & HR 8815 & GZ Peg      &   SRa  & 77 Jun 4     & 0.17 & --\tablenotemark{b} &    \\       
224935 & HR 9089 & YY Psc      &   Lb:  & 76 Jun 18    & 0.09 & 31                  &    \\     
260297 &         & DY Gem      &   SRa  &  87 Apr 14   & 0.07 & 93                  & 37 \\     
       & BD+48 1187 & TV Aur   &   SRb  & 77 Aug 25    & 0.08 & 56                  &    \\
       & BD+48 1187 & TV Aur   &   SRb  & 87 Apr 14    & 0.07 & 62     & 24          \\
\enddata
\tablenotetext{a}{Variable class from GCVS}
\tablenotetext{b}{Only one scan, no difference spectrum}
\end{deluxetable*}

\subsection{Observations}
A description of the FTS can be found in Paper I as well as in \citet{pilachowski_et_al_2017}.
In summary, the FTS was operated between the years 1975 and 1995 by KPNO as a coude instrument on the Mayall Telescope \citep{hall_et_al_1979}.
The FTS was a nearly perfect spectrograph for observing bright objects. 
Fourier spectrometers have very large wavelength coverage, no scattered light, and a well-defined instrumental profile. The broad wavelength coverage resulted in a highly useful archive.

The program spectra consist of  observations covering the 1.5 to 2.5\,$\mu$m ($H$ \& $K$) spectral region.   
The $H$ window covers the weaker CO $\Delta$v=3 lines while the long-wavelength side of the $K$ window
covers the CO $\Delta$v=2 bands.  This provides a range of line strengths for the common isotopologue.  
In the previous paper the observations were limited to those made with a single pair of detectors with matching filters.  
In this paper this has been broadened to include spectra utilizing a dichroics with spectra taken through separate $H$ and $K$ filters.  
The wavelength bounds of the spectra are the same, with the main difference being the details of the filter response.

The natural unit of an FTS is inverse wavelength, i.e. wavenumber ($\sigma$) and in standard convention these are cm$^{-1}$ units.  
The spectral resolution ($R=\sigma/\Delta\sigma$) is constant in wavenumber.   
High-resolution spectra taken with the 4 meter FTS typically have a resolution of $\sim$0.070 cm$^{-1}$.  
This corresponds to R$\sim$57000 at the red end (4000 cm$^{-1} = 2.5~\mu$m) of the spectrum and R$\sim$96000 at the blue end (6670 cm$^{-1} = 1.5~\mu$m).  
FTS spectra are typically apodized to dampen the sidelobes of the sinc function instrumental profile.  
Damping the amplitude of the instrumental profile sidelobes lowers the resolution and results in a corresponding increase in signal-to-noise.  
All spectra discussed here have been apodized by function I2 of \citet{norton_beer_1976}.

The FTS generated two interferograms for each observation.  
The resulting spectra are identical except for the noise.  
The S/N values appearing in Table \ref{t:obslog} are the RMS 
of the difference compared to the peak signal.  
For late-type stars the peak signal for observations covering the $K$ and $H$ band simultaneously always occurs in the $K$-band.  
For observations taken with the dichroic we present separate $H$- and $K$-band S/N.

\subsection{Telluric correction}

The technology used for the FTS was suitable for bright stars.  The high-resolution $K$-band limit was $\sim$ +4.  
Most of the stars in our sample are brighter.
The intrinsic colors of M stars have $K-V$ $>$ 3 \citep{tokunaga_2002}.
Since the intrinsic colors of hot stars have $K-V$ $\sim$ 0 it is clearly very difficult to find a suitably bright telluric reference
star for a bright M giant.  
As a result, most FTS spectra do not have an associated telluric reference star.  
Furthermore, the FTS spectra do not have meta-data providing information about the weather conditions.  
In the past we have not attempted any telluric correction and relied entirely on the large number of molecular lines present in the very large FTS bandpass.  
However, the rare oxygen CO isotopologues do not have that many available lines.  
The situation is tractable for $^{12}$C$^{17}$O since the 2-0 band has a conspicuous series of lines in spaces between telluric lines from 4280 to 4295 cm$^{-1}$.  
But no such region exists for $^{12}$C$^{18}$O.  
For this isotopologue we have to rely on isolated lines.   
In an attempt to reduce blending and increase the selection of lines, we have corrected the most important oxygen isotope regions for telluric absorption using synthetic telluric spectra. 
The suite of available options is discussed by \citet{seifahrt_et_al_2010}.  
We selected the ATRAN tool \citep{lord_1992} because of its simplicity. 
Our technique is to estimate airmass and precipitable water from the 4511 - 4516 cm$^{-1}$ region.  
This region has minimal stellar lines with isolated H$_2$O and $^{12}$CH$_4$ telluric lines.  
Based on these estimates for airmass and precipitable water we applied a telluric correction to the oxygen isotope regions. 
The correction is relatively crude, for instance, we have not attempted to modify the telluric atmosphere for temperature, but nonetheless the method satisfactorily removes unsaturated telluric lines.

\section{STELLAR PARAMETERS} \label{sect:parameters}

\subsection{Effective Temperatures and CO Excitation Temperatures} \label{sect:efftemp}

In Paper I isotopic abundances were derived using a simple curve-of-growth technique.
This is a robust technique for determining isotopic ratios.  It relies on a small number of 
assumptions by comparing lines of similar depths.  As the first step in the current analysis we have
employed this technique.
The excitation temperature of the weak CO lines, which directly affects the isotopic ratios, 
was extracted by forcing high- and low-excitation lines with the same observed equivalent width 
to the same location on the curve of growth \citep{hinkle_et_al_1976}. 
We used the same technique in Paper I to derive isotopic ratios from the spectra of mira stars. 
FTS spectra are perfectly suited for the application of this technique because of the large wavelength coverage.  In the
1.5 - 2.5 $\mu$m region of late-type stars the large number of CO lines observed 
results in a well-defined curve of growth.  This wavelength region spans CO first and second overtones 
and the large range in oscillator strength makes possible the measurement isotope ratios spanning three orders of magnitude.
For an illustration of this technique we refer to figure 2 of Paper I.  The technique does require that an excitation temperature
be assigned to the CO.
The accuracy with which the excitation temperature can be found naturally depends on the number of usable lines.
Typically, the excitation temperature could be constrained to better than $\pm$150 K. Fortunately, the isotopic ratio
is not highly sensitive to the excitation temperature and uncertainties in the $\pm$150 range have only a small effect on 
the resulting isotopic ratios.

\begin{figure}
\includegraphics[width=\columnwidth]{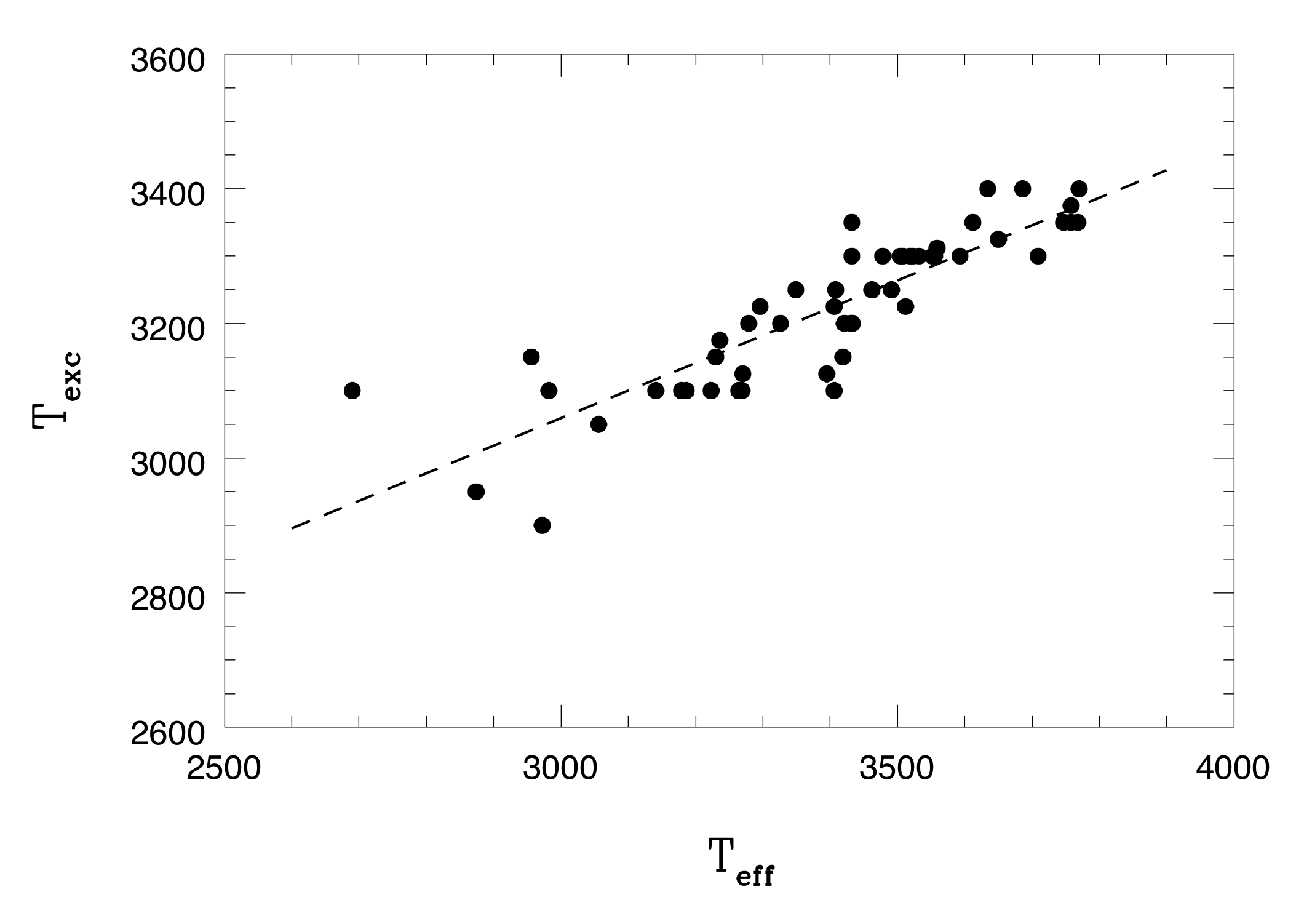}
\caption{
Excitation temperatures of the CO second-overtone lines plotted as a function of effective temperature.  
The effective temperatures for our M- and S-type stars are from V-K colors following \citet{richichi_et_al_1999}.  
Dashed line is a fit to the M and S relation. 
\label{f:teff_texc}
}
\end{figure}

To reduce the uncertainty in the abundances in the case of ill-defined excitation temperatures, we investigated the relation between the excitation temperature, $T_{\rm exc}$, and the effective temperature, $T_{\rm eff}$.  
There is considerable basis for assuming these quantities are related via the atmospheric structure and elemental abundances.  
Effective temperature is a measure of the total energy integrated over all wavelengths radiated from a unit of stellar surface area.
The value is fixed by luminosity and radius and is an observable quantity. 
$T_{\rm eff}$, is given by the relation
\begin{equation}
F = \left(\frac{\phi}{2}\right) ^2 \sigma T_{eff}^4
\end{equation}
where $F$ is the bolometric flux received at the Earth, $\sigma$ is the Stefan-Boltzmann constant, and $\phi$ is the angular diameter \citep{ridgway_et_al_1980}.  
Most values of $T_{\rm eff}$ for cool giants have been set using equation 1 with $\phi$ determined by lunar occultations and interferometry \citep[see, for instance,][]{ridgway_et_al_1980, richichi_et_al_1999} and $F$ determined from near-IR photometry.
Observational limits associated with the techniques restrict the sample of objects both in magnitude, diameter, and position on the sky.  
A second approach based on infrared flux \citep{blackwell_shallis_1977, blackwell_et_al_1990} also produces effective temperatures. 
Both techniques produce similar results.  
We limit our discussion to the angular diameter approach. 

Effective temperature is known to vary with spectral type, for the M giants nearly linearly \citep{richichi_et_al_1999}.
\citet{worthey_lee_2011} present a comparison of $T_{\rm eff}$ versus $V-K$ relations for cool giants.  
The \citet{bessell_et_al_1998} $T_{\rm eff}$ versus $V-K$ relation, derived from a compilation of occultation and interferometry diameters, is the best fit to the \citet{worthey_lee_2011} data.
\citet{sharma_et_al_2016} review a number of fits to the $T_{\rm eff}$ versus M giant spectral type relation against several groups of synthetic spectra.  
A $T_{\rm eff} - (V-K)$ relation  derived from a homogeneous set of angular diameters by \citet{richichi_et_al_1999} gives the best match and extends to the coolest non-Mira M giants. 
Therefore, we adopted the spectral type - effective temperature relation by \citet{richichi_et_al_1999} for our study. 
However, use of the \citet{bessell_et_al_1998} calibration would result in negligible change.   
All the calibrations show scatter around a mean relation.  
The \citet{richichi_et_al_1999} relation shows considerable scatter to lower temperatures.

Since the sample stars are bright, in fact most (Table \ref{t:obslog}) are in the Yale Bright Star Catalog \citep{hoffleit_1964}, there are well-determined spectral types and magnitudes.  
We ignore the galactic reddening of these nearby objects.
Table \ref{t:texc} lists the parameters for the program stars, including the effective temperatures derived using the \citet{richichi_et_al_1999} relation for the M giants.  
\citet{smith_lambert_1990} find that S stars of temperature subclass 5 and earlier follow the same effective temperature versus $V-K$ temperature subclass relation as for the M giants.
Following from this, we have assigned effective temperatures for the six S stars earlier than S5 in our data set by using the \citet{richichi_et_al_1999} scale.  
One star, DY Gem, is later than S5.  
We assume, as did \citet{smith_lambert_1990}, that the effective temperature of this star is the same as an M giant of the same subclass.

Using the method described in Paper I and in \citet{hinkle_et_al_1976}, we found empirical CO second-overtone excitation temperatures for the sample stars.  
The resulting excitation temperatures are listed in Table \ref{t:texc} and plotted as a function of effective temperature in Figure \ref{f:teff_texc}. 
A linear least-squares fit gives
\begin{equation}
T_{exc}~=~0.4091\,T_{eff}\,+\,1832.
\end{equation}
\citet{tsuji_1991} previously noted the relation between CO second-overtone excitation temperature and effective temperature.  

\startlongtable
\begin{deluxetable*}{llrrrlrrrrrr}
\tabletypesize{\footnotesize}
\tablecaption{Effective and excitation temperatures and luminosities for our sample stars  \label{t:texc}}
\tablewidth{0pt}
\tablehead{ \colhead{HD/BD} & \colhead{Spec.} & \colhead{V} & \colhead{K} & \colhead{V-K\tablenotemark{a}} &  
\colhead{Date} & \colhead{$T_{\rm eff}$} & \colhead{$T_{\rm exc}$} & \colhead{$T_{\rm exc}$} & \colhead{T$_{exc}$} & \colhead{log (L/L$_{\sun}$)} & \colhead{D\tablenotemark{b}}\\
\colhead{} & \colhead{Type\tablenotemark{a}} & \colhead{} &  \colhead{} & \colhead{} &
 \colhead{} & \colhead{K} & \colhead{(est.) K} & \colhead{(fit) K} & \colhead{(est.-fit)} & \colhead{} & \colhead{pc}
}
\startdata
HD   6860 &  M0III    & 2.05 & -1.87 & 3.92 & 76 Jun 6     & 3770 & 3400 & 3374 &   26 & \dots & \dots\\ 
HD   6860 &  M0III    &               &       &      & 79 Jun 14+15 &  & 3400  &       &   &  & \\  
HD   7351 &  S3+/2-   & 6.38 &  1.64 & 4.74 & 77 Jun 29    & 3532 & 3300 & 3277 &   23 & 3.132 & 282$^{-14}_{+16}$\\ 
HD  18191 & M6III     & 5.93 & -1.02 & 6.95 & 76 Aug 17    & 3236 & 3175 & 3156 &   19 & 3.178 & 98$^{-5}_{+6}$\\ 
HD  18884 &  M1.5IIIa & 2.53 & -1.67 & 4.20 & 76 Sep 28    & 3686 & 3400 & 3340 &   60 & \dots & \dots\\ 
HD  20720 &  M3/4III  & 3.70 & -1.15 & 4.85 & 76 Sep 29    & 3508 & 3300 & 3267 &   33 & 3.307 & 106$^{-6}_{+7}$\\ 
HD  22649 &  S3.5/2   & 5.11 &  0.10 & 5.01 & 76 May 14    & 3478 & 3300 & 3255 &   45 & 3.363 & 182$^{-9}_{+10}$\\ 
HD  30959 &  M3SIII   & 4.75 & -0.66 & 5.41 & 76 Aug 23    & 3421 & 3200 & 3232 &  -32 & 3.427 & 163$^{-11}_{+12}$\\ 
HD  30959 &  M3SIII   &               &       &      & 78 Jan 22    &       & 3200  &       &   &   & \\ 
HD  39225 & M1.5II-III& 6.07 &  1.94 & 4.13 & 89 Sep 14    & 3709 & 3300 & 3349 &  -49 & 2.736 & 190$^{-6}_{+7}$\\ 
HD  44478 &  M3IIIab  & 2.87 & -1.49 & 4.36 & 76 Sep 29    & 3634 & 3400 & 3319 &  81 & \dots & \dots\\  
HD  44478 &  M3IIIab  &               &       &      & 79 Sep 5+7+9 &       & 3300  &       &   &   & \\ 
HD  49368 &  S3/2     & 7.78 &  2.46 & 5.32 & 87 Apr 14    & 3432 & 3350 & 3236 &  114 & 3.501 & 754$^{-23}_{+25}$\\ 
HD  55966 &  M4IIIa   & 4.96 & -0.15 & 5.11 & 90 Apr 12    & 3462 & 3250 & 3248 &    2 & 3.256 & 154$^{-5}_{+5}$\\ 
HD  58521 &  M6S      & 6.98 & -0.44 & 7.42 & 78 Jan 22    & 3179 & 3100 & 3133 &  -33 & \dots & \dots\\ 
HD  64332 &  S4.5/2   & 7.64 &  2.31 & 5.33 & 87 Apr 13    & 3431 & 3200 & 3236 &  -36 & 3.541 & 627$^{-27}_{+30}$\\ 
HD  71250 &  M3III    & 5.50 &  0.56 & 4.94 & 78 Oct 15    & 3491 & 3250 & 3260 &  -10 & 3.576 & 303$^{-28}_{+34}$\\ 
HD  94705 &  M5.5III  & 5.78 & -0.81 & 6.59 & 90 Apr 11    & 3279 & 3200 & 3174 &   26 & 3.170 & 106$^{-5}_{+6}$\\                           
HD  96360 &  M3[Swk]  & 8.09 &  2.77 & 5.32 & 87 Apr 14    & 3432 & 3300 & 3236 &   64 & 2.868 & 417$^{-7}_{+7}$\\ 
HD 100029 &M0III-IIIa & 3.85 & -0.12 & 3.97 & 90 Apr 12    & 3758 & 3375 & 3369 &    6 & 3.139 & 119$^{-6}_{+7}$\\  
HD 102212 &  M1III    & 4.04 & 0.07  & 3.97 & 82 Apr 8     & 3758 & 3350 & 3369 & -19 & 2.915 & 100$^{-6}_{+7}$\\
HD 106198 & M4/5(I/II)& 6.47 & -0.24 & 6.71 & 87 Apr 13    & 3264 & 3100 & 3167 &  -67 & 3.907 & 431$^{-46}_{+59}$\\ 
HD 112300 &  M2III    & 3.38 & -1.24 & 4.62 & 76 Jun 18    & 3559 & 3312 & 3288 &   24 & \dots & \dots\\ 
HD 112300 &  M2III    &               &       &      & 77 Jun 24    &       & 3300  &       &   &   & \\ 
HD 114961 &  M7 III   & 6.85 & -1.79 & 8.64 & 76 Jun 18    & 3056 & 3050 & 3082 &  -32 & 4.506 & 316$^{-61}_{+100}$\\ 
HD 119228 &  M2III    & 4.66 &  0.35 & 4.31 & 90 Apr 11    & 3650 & 3325 & 3325 &    0 & 3.237 & 168$^{-8}_{+8}$\\ 
HD 121130 &  M3III    & 4.66 & -0.17 & 4.83 & 77 Jun 28    & 3512 & 3225 & 3269 &  -44 & 3.115 & 122$^{-4}_{+4}$\\  
HD 123657 &  M4.5 III & 5.28 & -0.37 & 5.65 & 76 Aug 18    & 3395 & 3125 & 3221 &  -96 & 3.537 & 190$^{-12}_{+14}$\\  
HD 126327 &  M7-8IIIe & 8.60 & -1.96 &10.56 & 76 Jan 14    & 2874 & 2950 & 3008 &  -58 & 3.633 & 128$^{-5}_{+5}$\\ 
HD 132813 &  M5III    & 4.54 & -1.00 & 5.54 & 76 Aug 18    & 3406 & 3100 & 3225 & -125 & 3.176 & 100$^{-6}_{+7}$\\ 
HD 133216 &  M3/4III  & 3.21 & -1.43 & 4.64 & 76 Aug 18    & 3555 & 3300 & 3286 &   14 & 3.587 & 128$^{-12}_{+15}$\\ 
HD 146051 &  M0.5III  & 2.75 & -1.26 & 4.01 & 77 Jun 24    & 3747 & 3350 & 3365 &  -15 & \dots & \dots\\  
HD 148783 &  M6-III   & 5.01  & -2.05 & 7.06 & 77 Jun 30    & 3223 & 3100 & 3151 &  -51 & 3.696 & 122$^{-5}_{+6}$\\ 
HD 149683 &  M7III:e  & 9.44  & -0.04 & 9.48 & 83 Sep 13    & 2982 & 3100 & 3052 &   48 & 3.786 & 384$^{-16}_{+18}$\\                           
HD 154143 &  M3III    & 4.98 &  0.49 & 4.49 & 90 Apr 12    & 3593 & 3300 & 3302 &   -2 & 2.913 & 126$^{-3}_{+3}$\\                           
HD 163990 & M6S       & 6.32 &  0.07 & 6.25 & 76 Sep 29    & 3326 & 3200 & 3193 &   7 & 3.676 & 275$^{-13}_{+15}$\\ 
HD 168574 & M3III     & 6.25 & -0.21 & 6.46 & 76 Sep 29    & 3296 & 3225 & 3180 &   45 & 3.825 & 322$^{-23}_{+26}$\\ 
HD 175588 & M4II      & 4.30 & -1.22 & 5.52 & 76 Apr 11    & 3408 & 3250 & 3226 &   24 & \dots & \dots\\ 
HD 175865 & M5III     & 4.00 & -2.08 & 6.08 & 90 Apr 11    & 3349 & 3250 & 3202 &   48 & 2.509 & 96$^{-5}_{+6}$\\ 
HD 182917 & M7IIIab   & 7.08 & -0.70 & 7.78 & 79 Feb 8     & 3141 & 3100 & 3117 &   -17 & 3.302 & 183$^{-7}_{+8}$\\                           
HD 183439 &  M0III    & 4.45 &  0.52 & 3.93 & 76 May 16    & 3768 & 3350 & 3374 &  -24 & 2.744 & 100$^{-3}_{+3}$\\  
HD 184313 & M5-5.5III & 6.48 & -0.18 & 6.66 & 84 Feb 21    & 3270 & 3125 & 3170 &  -45 & 3.508 & 218$^{-10}_{+11}$\\                           
HD 184313 & M5-5.5III &               &       &      & 84 Nov 13    &       & 3100  &       &    &   &\\     
HD 196610 & M6III     & 5.89 & -1.11 & 7.00 & 90 Apr 12    & 3230 & 3150 & 3153 &    -3 & 3.291 & 110$^{-4}_{+4}$\\                           
HD 198026 & M3III     & 4.44 & -0.21 & 4.65 & 76 May 16    & 3552 & 3300 & 3285 &   15 & 3.161 & 126$^{-5}_{+6}$\\ 
HD 198164 & SC2/7.5   &11.00 &  2.18 & 8.82 & 76 Jun 19    & 2690 & 3100 & 2933 &  167 & 4.213 & 1509$^{-77}_{+86}$\\                           
HD 200527 & M3Ib-II   & 6.23  &  0.92 & 5.31 & 87 Apr 13    & 3433 & 3200 & 3237 &  -37 & 3.674 & 391$^{-21}_{+24}$\\ 
HD 202369 & M2III     & 5.34    &  0.56 & 4.78 & 76 May 12    & 3523 & 3300 & 3273 &   27 & 3.156 & 173$^{-9}_{+10}$\\ 
HD 205730 & M4e-6eIII & 5.38 & -1.29 & 6.67 & 77 Apr 4     & 3269 & 3100 & 3169 &  -69 & \dots & \dots\\ 
HD 205730 & M4e-6eIII &               &       &      & 84 Apr 8     &       & 3000  &       &   &  & \\   
HD 209872 & M7        & 9.20 & -0.55 & 9.75 & 83 May 21    & 2956 & 3150 & 3041 &  109 & \dots & \dots\\                           
HD 209872 & M7        &               &       &      & 84 Nov 13    &       & 3150  &       &   &   & \\    
HD 216386 & M1.2III   & 3.79 & -0.64 & 4.43 & 76 Jun 6     & 3612 & 3350 & 3310 &   40 & 3.090 & 94$^{-5}_{+5}$\\ 
HD 216672 & S4+/1+    & 6.36 &  0.94 & 5.42 & 77 Jun 4     & 3419 & 3150 & 3231 &  -81 & 3.943 & 422$^{-28}_{+32}$\\                           
HD 216672 & S4+/1+      &               &       &      & 88 Jul 2 & & 3200 &       &   &  & \\
HD 217906 &M2.5II-III & 2.42 & -2.38 & 4.80 & 76 May 12    & 3519 & 3300 & 3272 &   28 & \dots & \dots\\ 
HD 218634 & M4S       & 5.14 & -0.40 & 5.54 & 77 Jun 4     & 3406 & 3225 & 3225 &    0 & 3.719 & 219$^{-22}_{+27}$\\                           
HD 224935 & M2III     & 4.41 & -0.46 & 4.87 & 76 Jun 18    & 3504 & 3300 & 3266 &   34 & 3.255 & 127$^{-7}_{+8}$\\ 
HD 260297 & S8,5      &11.20  &  1.62 & 9.58 & 87 Apr 14    & 2972 & 2900 & 3048 & -148 & \dots & \dots\\                           
BD+48 1187& S5/6      & 9.32  &  1.96 & 7.36 & 77 Aug 25    & 3186 & 3100 & 3135 &  -35 & 4.064 & 616$^{-41}_{+48}$\\                           
\enddata
\tablenotetext{a}{Spectral type, V, K from Simbad.}
\tablenotetext{b}{Gaia distances and uncertainties from \citet{bailer_jones_et_al_2018}}
\end{deluxetable*}

\subsection{Surface Gravity}

For curve-of-growth analysis only the excitation temperature is required.  However, since the 
stars in the sample have only small velocity amplitudes, it seemed reasonable to explore deriving 
abundances using spectrum synthesis.  This requires the use of a stellar model atmosphere where the inputs
include surface gravity.  To determine the surface gravity 
we extracted the distances and brightnesses of the sample stars from \citet{bailer_jones_et_al_2018} based on the Gaia DR2 catalog \citep{gaia_dr2}.  We were able to attribute distances to 40 of the 51 stars in the sample. 
With these distances we derived luminosities with the help of the mean Gaia G, BP, and RP magnitudes and bolometric corrections given in \citet{lebzelter_et_al_2019} which are based on a library of reference objects from \citet{kerschbaum_et_al_2010}.  Limitations in the astrometric solution provided in Gaia DR2\footnote{https://www.cosmos.esa.int/web/gaia/dr2} could affect the luminosity values used here.
Using the luminosity and effective temperature from equation 2 the radius follows from equation 1.  As a starting point, we assume a mass of 
1.2\,M$_{\sun}$ for a typical AGB star.  These values were in turn used to select model atmospheres from a grid of atmospheres (see Section \ref{sect:synth}.  The grid is set in steps of log g.  Since the entire range of AGB masses covers less than 1 in the log, the assumed value of 1.2 \,M$_{\sun}$ does not have a large impact on the model atmosphere selected.  

\section{ABUNDANCES}\label{sect:abundances}

\subsection{Isotopic ratios of Carbon and Oxygen}
Isotopic ratios were determined using the technique described in Paper I.  The technique compares weak lines of equal strengths from different isotopes.  The isotopic ratio is the shift in the abscissa between the curve of growth for the $^{12}$C$^{16}$O lines and the curve of growth for the rarer isotope.
The LTE line strength is plotted on the abscissa, and the reduced equivalent width (equivalent width divided by the wavenumber of the transition) is plotted on the ordinate. 
References to the input data on the transitions used in our analysis can be found in Paper I.

As in Paper I, the set of lines was selected by searching each vibration-rotation transition that falls in the $H$ and $K$ bands.  For $^{12}$C$^{17}$O, and $^{12}$C$^{18}$O the selection is limited to lines from only 2-0 and 3-1.  The $^{12}$C$^{17}$O 2-0 lines from R20 to R33 are well placed \citep[see figure 6 of][]{wallace_hinkle_1996} and unless the ratio $^{16}$O/$^{17}$O ratio is very large, lines can be found.  On the other hand, the  $^{12}$C$^{18}$O lines are mixed in with other CO lines.  Nonetheless, a few fairly convincing lines were typically found in each spectrum.  A list of lines that show little blending is given in Paper I. 
For $^{12}$C$^{16}$O and $^{13}$C$^{16}$O lines from the various first- and second-overtone bands were used.
Central depths were measured for these lines and transferred into reduced equivalent width.

For all the program stars lines were measured in both the first ($\Delta$v=2) and second ($\Delta$v=3) $^{12}$C$^{16}$O overtones.  For $^{13}$C$^{16}$O lines were always present from the first overtone and for cooler stars from both overtones.  The two overtones were combined into a single curve of growth 
for each isotope.  The shifts between the curves of growth of the isotopologue give the isotopic ratios.  
The extreme range of possible shifts is shown by the uncertainties.
The resulting isotopic ratios and their uncertainties can be found in Table  \ref{t:meas_values}.
For some stars we also explored varying the excitation temperature.
This was done both to evaluate the dependency of the results on this input parameter and to handle cases where the empirical determination gave a strongly deviating result from the $T_{\rm exc}-T_{\rm eff}$ relation derived in Section \ref{sect:efftemp}.  Paper I demonstrated that the error introduced by typical 100 K uncertainties in the excitation temperature corresponds to errors in the isotopic ratios of C and O of $\le$10\,\%.  

Where we had more than one spectrum for a star, the differences are typically within the uncertainty limits.
Not surprisingly, a lower S/N can prevent the determination of the isotopic ratios of oxygen, for instance in $\delta$ Vir and W Cyg.
The oxygen isotope ratios always compare the $K$ band first-overtone lines against $^{12}$C$^{16}$O second-overtone lines in the 
$H$ band.  The S/N typically peaks in the middle of the $K$ band and is about a factor of two lower in the $H$ band.  Since lines of equal strength 
are compared, in spectra with marginal peak S/N the weak rare oxygen isotopic lines are then compared to weak $^{12}$C$^{16}$O dominated 
by noise.  
This may also be the cause for the very large difference in $^{16}$O/$^{18}$O observed between the two observations of $\mu$ Gem.
Note that the other isotopic ratios for this star are in reasonable agreement.
In Paper I we showed that the pulsation phase has very little influence on the measured isotopic ratio.
For our sample consisting of small-amplitude variables we expect this effect to be even less relevant.

The shift along the abscissa that gives the isotopic ratio is on a logarithmic scale.
When the standard deviation of the shift is then transferred into a linear scale, the resulting error is asymmetric.
In some earlier papers, also in our Paper I, a single mean value was given for the error, overestimating the uncertainty on the negative side and underestimating it on the positive side.
In the present paper we decided to list the asymmetric error and to use it also in the plots where applicable.

\startlongtable
\begin{deluxetable*}{llllllll}
\tabletypesize{\footnotesize}
\tablecaption{Measured Values for C and O Isotopes}  \label{t:meas_values}
\tablewidth{0pt}
\tablehead{
\colhead{HD/BD} &
\colhead{Var} & 
\multicolumn{2}{c}{$^{12}$C/$^{13}$C} & 
\multicolumn{2}{c}{$^{16}$O/$^{17}$O} &
\multicolumn{2}{c}{$^{16}$O/$^{18}$O} \\
\colhead{} &
\colhead{name} &
\colhead{mean} &
\colhead{low/hi} &
\colhead{mean} &
\colhead{low/hi} &
\colhead{mean} &
\colhead{low/hi}
}
\startdata
HD 6860 & $\beta$ And 06/76   & 10        & 9/11    & 239    & 181/316   & 2199   & 1825/2649 \\
HD 6860 & $\beta$ And 06/79  & 10       & 9/12    & 158    & 122/204   & 578    & 240/1389 \\
HD 7351 & DT Psc        & 16        & 13/19    & $>$5000  & \dots         & \dots      &  \\
HD 18191 & RZ Ari       & 6       & 5/7    & 958    & 491/1870  & 2422   & 1626/3607\\
HD 18884 & $\alpha$ Cet & 11        & 6/20    & 660    & 532/819   & 730    & 590/904\\
HD 20720 & $\tau^{4}$ Eri  & 8     & 7/11    & 598    & 189/1891  & 473    & 181/1238\\
HD 22649 & BD Cam   & 14       & 10/20    & 486    & 269/880   & 140    & 77/255\\
HD 30959 & o$^{1}$ Ori 08/76 & 16   & 14/21    & 532    & 397/713   & 513    & 260/1012\\
HD 30959 & o$^{1}$ Ori  01/78 & 15    & 14/19    & 360    & 306/424   & 827    & 438/561\\
HD 39225 & HR 2028  & 10      & 7/13    & 600    & 206/1750  & 977    & 482/1981\\
HD 44478 & $\mu$ Gem 09/76 &  11        & 5/23    & 618    & 347/1101  & 2430   & 1444/4087\\
HD 44478 & $\mu$ Gem  09/79 & 10       & 8/13    & 660    & 445/978   & 409    & 193/868\\
HD 49368 & V613 Mon    & 16     & 13/20    & 556    & 333/928   & 1103   & 663/1835\\
HD 55966 & VZ Cam   & 14      & 11/18    & 3924   & 2942/5235 & 761    & 514/1126\\
HD 58521 & Y Lyn   & 39   & 32/47   & 1087   & 479/2465  & 387    & 233/643\\
HD 64332 & NQ Pup  & 20      & 18/22    & 534    & 319/893   & 1643   & 744/3628\\           
HD 71250 & BP Cnc   & 9       & 7/11    & 3284   & 1786/6041 & 2064   & 1130/3773\\
HD 94705 & VY Leo   & 8       & 6/11     & 2187   & 1309/3655 & 2418   & 1279/4571\\
HD 96360 & HL UMa   & 11  & 8/16    & $>$5000  & \dots & 355    & 202/623\\
HD 100029 & $\lambda$ Dra  & 9     & 7/13    & 1290   & 705/2361  & 664    & 262/1683\\   
HD 102212 & $\nu$ Vir &  8 & 6/10 & $>$2300 & \dots & $>$2300 & \dots\\
HD 106198 & V335 Hya  & 18       & 14/20    & 497    & 265/934   & 562    & 312/1015\\
HD 112300 & $\delta$ Vir 06/76  & 13      & 13/14    & \dots      & \dots         & \dots      & \dots    \\
HD 112300 & $\delta$ Vir 06/77  & 18        & 14/20    & $>$5000  & \dots & 928    & 568/1517\\
HD 114961 & SW Vir   & 30     & 22/40    & 530    & 173/1623  & 591    & 250/1400\\
HD 119228 & IQ UMa   & 14       & 13/18    & 1535   & 869/2713  & 657    & 166/2606\\
HD 121130 & CU Dra   & 10       & 8/15    & 228    & 177/295   & \dots      & \dots   \\
HD 123657 & BY Boo  & 10        & 9/13    & 2056   & 1586/2666 & \dots      & \dots   \\
HD 126327 & RX Boo   & 16        & 13/20    & 301    & 151/599   & 247    & 155/393 \\
HD 132813 & RR UMi  & 15       & 10/18    & $>$5000  & \dots & 656    & 434/992 \\
HD 133216 & $\sigma$ Lib  &  6         & 3/10    & $>$2000   & \dots         & $>$2000      & \dots  \\
HD 146051 & $\delta$ Oph  & 11      & 8/15    & 1424   & 895/2268  & 1322   & 414/4219 \\
HD 148783 & g Her   & 13        & 9/17    & 723    & 449/1166  & 853    & 412/1766 \\
HD 149683 & R UMi   & 14  & 11/19    & 1128   & 426/2982  & 618    & 545/702 \\
HD 154143 & HR 6337  & 12   & 10/15    & 1836   & 1097/3071 & 640    & 450/911 \\
HD 163990 & OP Her  & 14       & 10/18    & 435    & 230/821   & 478    & 244/939 \\
HD 168574 & V4028 Sgr   & 15  & 13/18    & 742    & 486/1134  & 1319   & 568/3065 \\
HD 175588 & $\delta^{2}$ Lyr  & 21    & 18/25    & 1035   & 830/1292  & 660    & 397/1098 \\
HD 175865 & R Lyr   & 18  & 16/21    & 829    & 400/1716  & 523    & 310/883 \\
HD 182917 & CH Cyg   & 15        & 12/18           & 885    & 490/1595  & 464    & 333/648 \\
HD 183439 & $\alpha$ Vul  & 14    & 10/17    & 2079   & 1096/3943 & 1880   & 522/6775 \\
HD 184313 & V450 Aql 02/84 & 14      & 11/17    & 964    & 348/2671  & 329    & 215/503 \\
HD 184313 & V450 Aql 11/84 & 14     & 12/17    & 1970   & 969/4004  & 431    & 210/882 \\
HD 196610 & EU Del   & 17   & 13/20    & $>$5000  & \dots & 1072   & 593/1939 \\
HD 198026 & EN Aqr  & 10       & 8/12    & \dots      & \dots         & \dots      & \dots   \\
HD 198164 & CY Cyg  & 7  & 6/11     & 94     & 25/342    & \dots      & \dots   \\
HD 200527 & V1981 Cyg  &  22        & 19/27   & 845    & 420/1699  & 745    & 297/1869 \\
HD 202369 & 29 Cap   & 12        & 9/14    & 1788   & 1412/2266 & \dots      & \dots   \\
HD 205730 & W Cyg 04/77  & 13       & 11/17    & 1448   & 964/2174  & \dots      & \dots   \\
HD 205730 & W Cyg 08/84  & 15        & 12/20    & 1713   & 905/3245  & 515    & 230/1154 \\
HD 209872 & SV Peg 05/83 & 20   & 17/23    & 168    & 12/2353   & 345    & 285/418 \\
HD 209872 & SV Peg 11/84 & 17  & 14/21    & 314    & 191/516   & 341    & 210/554 \\
HD 216386 & $\lambda$ Aqr & 9 & 4/15     & 4003   & 3748/4276 & 3167   & 1923/5215 \\
HD 216672 & HR Peg 06/76 & 22   & 18/28    & 737    & 492/1105  & 223    & 153/325 \\
HD 216672 & HR Peg 06/77  & 14 & 6/32 & 1800: & \dots & 2386 & 2106/2705 \\
HD 217906 & $\beta$ Peg & 9      & 8/10    & $>$5000  & \dots & 827    & 587/1163 \\
HD 218634 & GZ Peg  & 11 & 10/12    & 169    & 54/532    & 86     & 43/171  \\
HD 224935 & YY Psc  & 11 & 10/12    & 648    & 348/1205  & 450    & 243/833 \\
HD 260297 & DY Gem   & 56 & 40/79   & 733    & 285/1885  & 704    & 395/1257 \\
BD+48 1187 & TV Aur 08/77 & 29 & 26/32   & 213    & 82/552    & 149  & 68/325 \\
BD+48 1187 & TV Aur 04/87 & 52 & 43/63   & 237   & 194/291   & 342  & 315/371\\
\enddata
\end{deluxetable*}

\subsection{Comparison with literature values} \label{sect:literature}

\begin{figure}
    \centering
    \includegraphics[width=\columnwidth]{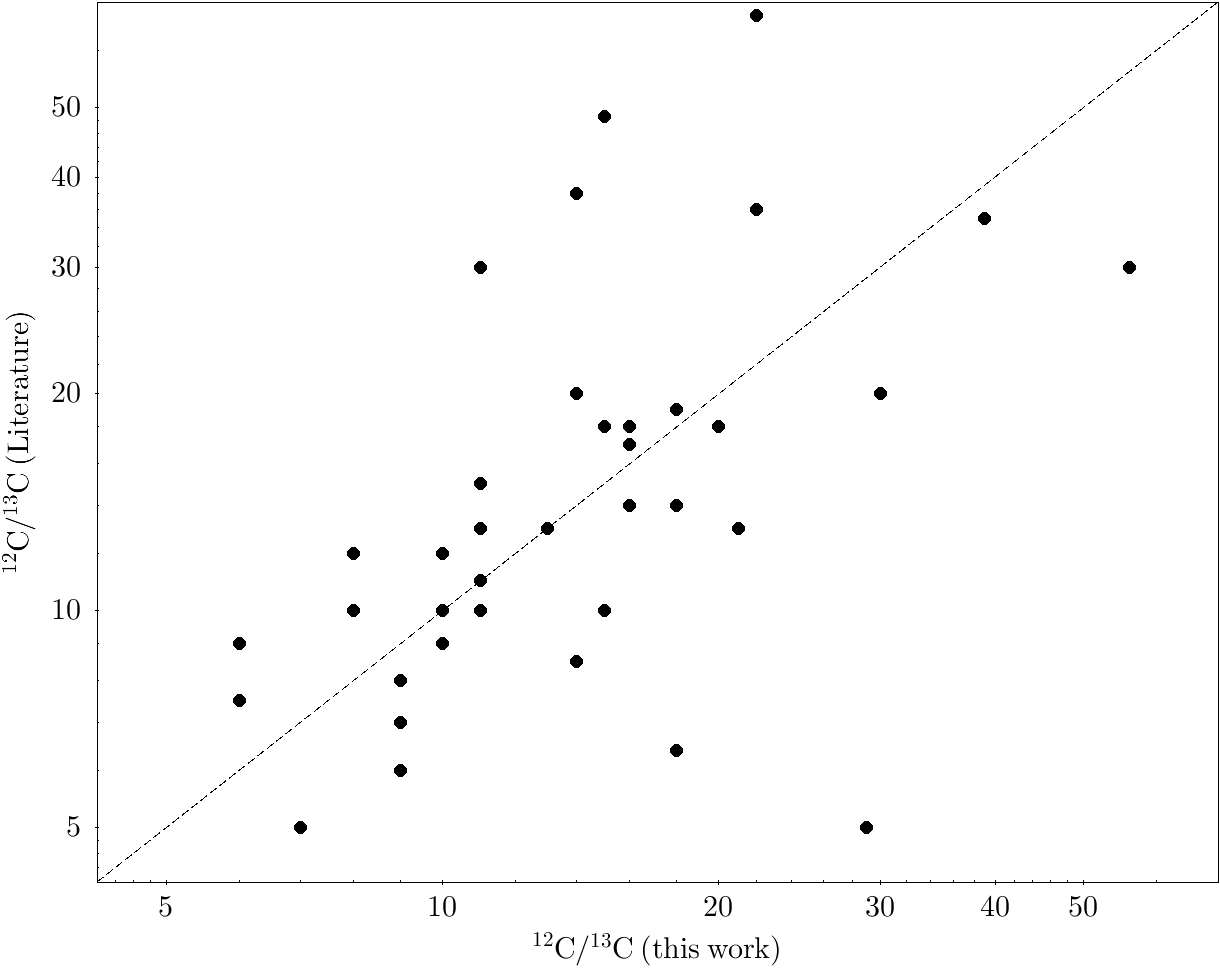}
    \caption{Comparison of $^{12}$C/$^{13}$C ratios from the literature with our results. The dashed line marks a 1:1 relation.}
    \label{fig:literatureCC}
\end{figure}

\begin{figure}
    \centering
    \includegraphics[width=\columnwidth]{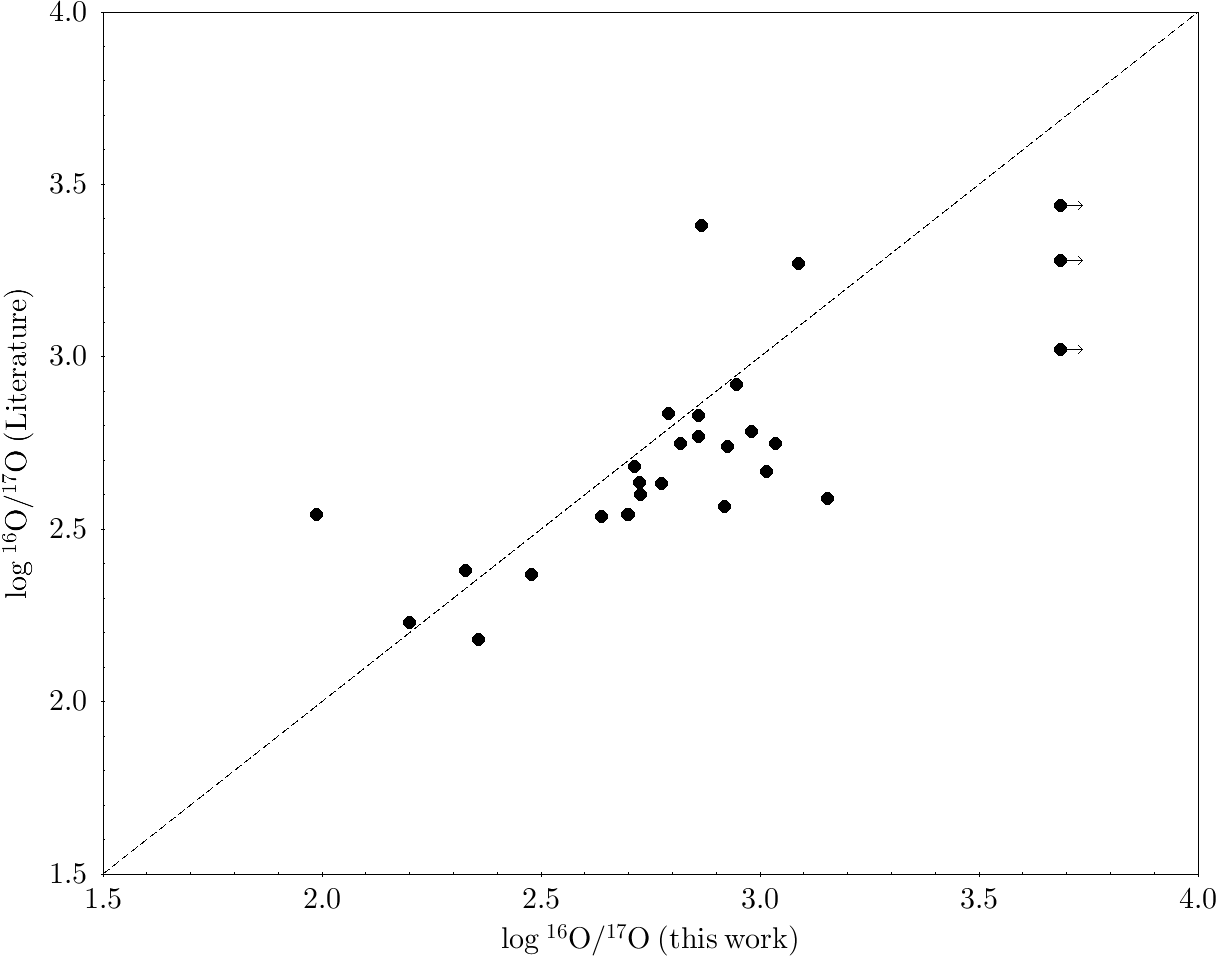}
    \caption{
    Comparison of $^{16}$O/$^{17}$O ratios from the literature with our results. The dashed line marks a 1:1 relation. Due to the large range covered by the isotopic ratios of oxygen, a logarithmic scale was used. Data points with arrows denote lower limits.}
    \label{fig:literatureOO}
\end{figure}

Various of the C and O isotope ratios for a number of the stars in the sample have previously been 
reported in the literature (Table  \ref{t:lit_values}).  Some of these results are based on the same spectra used here.                                        
References are given at the end of the table.
In some cases, the literature values for the same star, but not necessarily from the same data material, deviate significantly from each other.
For a comparison of previously determined values with our results (Figs.\,\ref{fig:literatureCC} and \ref{fig:literatureOO}) we averaged the isotopic ratios from the literature if they showed the same order of magnitude.
In the case values deviating more strongly, computing such an average would not have much value, and we decided to include the ratio closer to our result in the comparison.
Of course, this includes the risk of reinforcing a wrong result.

\startlongtable
\begin{deluxetable*}{lllll}

\tabletypesize{\footnotesize}
\tablecaption{Literature Values for C and O Isotopes  \label{t:lit_values}}
\tablewidth{0pt}
\tablehead{ \colhead{HD/BD} & \colhead{Var}  & \colhead{Lit $^{12}$C/$^{13}$C} & \colhead{Lit $^{16}$O/$^{17}$O} & \colhead{Lit $^{16}$O/$^{18}$O} \\
            \colhead{}      & \colhead{Name}     & \colhead{ ref (value)}           & \colhead{ ref (value)}          & \colhead{ ref (value)} 
}
\startdata
 HD 6860  &$\beta$ And&	1 (10), 4 (11$\pm$3)                           & 13 (170$\pm$10)                      & 13 (425$^{-75}_{+150}$)                   \\
 HD 7351  & DT Psc   &	4 (12$\pm$4), 12 (24$^{-8}_{+16}$)             & 12 (3000$^{-1200}_{+2500}$), 4 (2500) & 4 (2250), 12 (4600$^{-2100}_{+5900(!)}$) \\
HD 18191  & RZ Ari   &	1 (10), 11 (7.9$\pm$0.8)                       & 11 (607$\pm$48)                      & \dots                                    \\
HD 18884  &$\alpha$ Cet & 1 (10), 11 (11.1$\pm$0.8)                     & 11 (586$\pm$47)                      & \dots                                    \\
HD 20720  & $\tau^{4}$ Eri & 11 (12.4$\pm$0.3)                              & 11 (687$\pm$14)                      & \dots                                    \\
HD 22649  & BD Cam   & 2 (34), 4 (25$\pm$4), 12 (55$^{-18}_{+32}$)    &  4 (350), 12 (2250$^{-550}_{+700}$) & 12 (4700$^{-1400}_{+2200}$), 4 (650)     \\
HD 30959  & o$^1$ Ori& 1 (10), 4 (18$\pm$2), 12 (25$^{-6}_{+12}$)      &  4 (480), 12 (925$^{-225}_{+270}$) & 4 (640), 12 (1350$^{-450}_{+850}$)       \\ 
HD 39225  & HR 2028  &  3 (9$\pm$1.4)                                 &  3 ($>$260)                         & \dots                                    \\
HD 44478  & $\mu$ Gem &	2 (15), 4 (13$\pm$2), 11 (10.5$\pm$1.2)        &  11 (798$\pm$73), 13 (325$^{-75}_{+150}$) & 13 (475$^{-125}_{+200}$)             \\
HD 49368 & V613 Mon  &  4 (14$\pm$2)                                  & 4 (430)                             & 4 ($>$800)                                \\
HD 58521 & Y Lyn     &  4 (43$\pm$8), 10 (27)                         & 4 (560)                             & 4 (630)                                 \\
HD 64332 & NQ Pup    &  4 (18$\pm$4)                                  & 4 (400)                             & 4 (720)                                 \\
HD 96360 & HL UMa    &  4 (13$\pm$2)                                  & 4 (1900)                            &  4 (450)                                \\
HD 100029 &$\lambda$ Dra&  3 (7$\pm$1.7)                              & 3 ($>$85)                            & \dots                                    \\
HD 102212 & $\nu$ Vir & 4 (12$\pm$2), 11 (8.7$\pm$1.3) & 11 ($>$2000) & \dots \\
HD 106198 & V335 Hya &  4 (19$\pm$5)                                   & 4 (350)                             & 4 (1120)                               \\
HD 112300 &$\delta$ Vir& 4 (16$\pm$4), 11 (12.3$\pm$1.2)                 & 11 ($>$2500)                          & \dots                                    \\ 
HD 114961 & SW Vir   & 10 (18), 11 (22.0$\pm$4.7)                      & 11 (432$\pm$37)                      & \dots                                    \\
HD 121130 & CU Dra   &	2 (10), 4 (12$\pm$3),                         &  11 (151$\pm$11)                  & \dots                                    \\
         &           &  11 (14.8$\pm$1.6)                              &                                      &                                          \\
HD 123657 & BY Boo   &  4 (9$\pm$2)                                   & \dots                                & \dots                                    \\
HD 126327 & RX Boo   & 9 (17)                                        &  11 (233$\pm$20)                     & \dots                                    \\
HD 132813 & RR UMi   & 11 (10.0$\pm$0.8)                               & 11 ($>$2000)                          & \dots                                    \\
HD 133216 & $\sigma$ Lib &   11 (7.5$\pm$0.3)                           & 11 ($>$1500)                          & \dots                                    \\
HD 146051 &$\delta$ Oph & 1 (30), 11 (11.1$\pm$0.9)                     & 11 (387$\pm$68)                      & \dots                                    \\
HD 148783 & g Her    & 4 (10$\pm$2), 13 (12.5$\pm$1.1),                &  11 (211$\pm$42), 14 (675$\pm$175) & 12 (850$^{-300}_{+500}$)                    \\
         &           &                   12 (16$^{-6}_{+9}$)           &                                      &                                          \\
HD 149683 & R UMi    & 10 (15$\pm$3)                                   & 10 (1862$\pm$700)                   & 10 (525$\pm$120)                          \\
HD 163990 &OP Her  &  4 (20$\pm$2), 13 (11.3$\pm$1.2),             & 4 (360), 11 (329$\pm$31),            & 4 (590)                                \\
        &           &   12 (21$^{-7}_{+14}$)                          & 12 (850$^{-200}_{+275}$)             &  12 (1200$^{-350}_{+525}$)               \\
HD 168574 & V4028 Sgr &	11 (48.5$\pm$2.9)                              &  11 ($>$1000)                         & \dots                                    \\
HD 175588 & $\delta^2$ Lyr & 1 (10), 11 (16.2$\pm$1.5)                  & 11 (465$\pm$41)                      & \dots                                    \\
HD 175865 & R Lyr    & 11 (6.4$\pm$0.3)                                & 11 (368$\pm$44)                      & \dots                                    \\
HD 182917 & CH Cyg   & 14 (18$^{+12}_{-6}$)                           & 14 (830$^{+400}_{-270}$)             & 14 ($>$1000)                             \\
HD 183439 &$\alpha$ Vul & 1 (10), 4 (7$\pm$3)                          & \dots                                & \dots                                    \\
HD 198026 & EN Aqr   &	1 (10)                                        & \dots                                & \dots                                    \\
HD 198164 & CY Cyg   &  15 (5.6$\pm$1.2), 18 (5)                     &  15 (350$\pm$120), 19 (620$\pm$50)    & 16 (770$\pm$500)                         \\
HD 200527 & V1981 Cyg&	2 (46), 4 (23$\pm$5),                        & 4 (550), 12 (1850$^{-425}_{+550}$)  & 4 ($>$1300), 12 (3200$^{-900}_{+1250}$) \\
         &           &      12 (39$^{-12}_{+20}$)                      &                                      &                                          \\
HD 216386  &$\lambda$ Aqr & 1 (3), 11 (7.9$\pm$1.4)                     & 11 ($>$1000)                          & \dots                                    \\
HD 216672  & HR Peg &   2 (63), 12 (71$^{-21}_{+34}$)                &   12 (2400$^{-600}_{+850}$)          & 12 (3000$^{-1100}_{+2000}$)            \\
HD 217906  &$\beta$ Peg & 1 (10), 4 (8$\pm$2), 11 (7.7$\pm$0.5)      &  11 ($>$2500), 13 (1050$^{-250}_{+500}$) & 13 (425$^{-75}_{+150}$)                  \\
HD 218634  & GZ Peg  & 1 (30)                                         & \dots                                & \dots                                    \\
HD 224935  & YY Psc  & 1 (10)                                         & \dots                                & \dots                                    \\
HD 260297  & DY Gem  & 4 (30$\pm$7)                                   &  4 (570)                            & 4 (1210)                                \\
BD+48 1187 &TV Aur   & 4 (5$\pm$2)                                    & 4 (240)                             & 4 (440)                                 \\
\enddata

\tablecomments{Note that the references cover the period from 1985 to later. 
1=\citet{lazaro_et_al_1991} models only have 12/13 = 3, 10, 30, 60;
2=\citet{smith_lambert_1986}; 
3=\citet{wallerstein_morell_1994};
4=\citet{smith_lambert_1990} see Smith and Lambert 1985 ApJ 294 326 for stellar parameters;  
5=\citet{lambert_et_al_1986} $^{12}$C/$^{13}$C from table 4, uncertainties rough average of table 6, spectra same as ours;
6=\citet{ohnaka_tsuji_1996}; 
7=\citet{harris_et_al_1987};
8=\citet{schoier_olofsson_2000};
9=\citet{ramstedt_olofsson_2014};
10=\citet{hinkle_et_al_2016}; 
11=\citet{tsuji_2008};
12=\citet{harris_et_al_1985}; 
13=\citet{harris_lambert_1984};
14=\citet{schmidt_et_al_2006};
15=\citet{dominy_et_al_1986};
16=\citet{abia_et_al_2017}, spectra same as ours.
}
\end{deluxetable*}

It can be seen from Figure \ref{fig:literatureCC} that for the carbon isotopic ratio the majority of our results are in reasonable agreement with the literature results considering the uncertainties. 
We note that on average our values tend to be slightly lower than the results in the literature.
However, there are a few stars with strongly deviating $^{12}$C/$^{13}$C given in the literature, even though some results have been deduced from the spectra we used in our analysis.

\subsubsection{Discrepant $^{\it 12}$C/$^{\it 13}$C Cases}

HR Peg has been studied by \citet{smith_lambert_1986} who found $^{12}$C/$^{13}$C = 63, while our value is 22. 
A similar case, with the same literature reference, is BD Cam, for which we found $^{12}$C/$^{13}$C = 14 
while \citet{smith_lambert_1986} report a ratio of 34, later refined to 25 \citep{smith_lambert_1990}.  To 
make this more puzzling, the same archival FTS spectra were used in both analyses.  
The spectra have the problem of poor overlap between the $^{12}$CO and $^{13}$CO curves of growth. 
The $^{13}$CO second-overtone lines have depths of only a few percent of the continuum, and at a S/N around 50 in the $H$-band the noise is of similar size.
The S stars do have strong CO, and saturation is a problem for the first overtone.    

\begin{figure}
    \centering
    \includegraphics[width=\columnwidth]{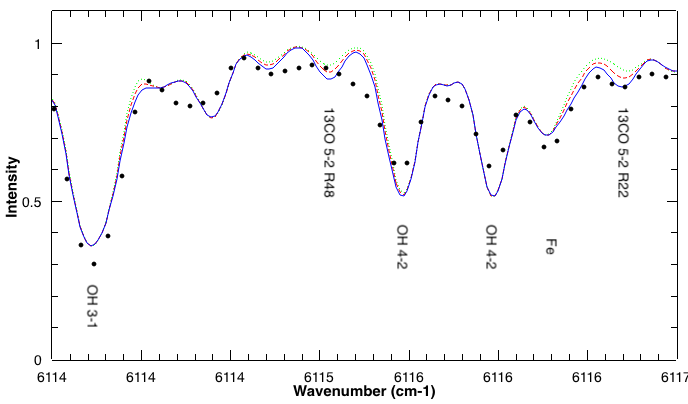}
    \caption{FTS spectrum of HR Peg compared with a hydrostatic model atmosphere with $T_{\rm eff}$=3500\,K and log\,g = 1.0. The solid (blue), dashed (red) and dotted (green) lines are for $^{12}$C/$^{13}$C ratios of 14, 25, and 62, respectively.}
    \label{fig:hrpeg}
\end{figure}

To explore this further a synthetic $H$-band spectrum was computed for HR Peg for different $^{12}$C/$^{13}$C ratios.
In Figure \ref{fig:hrpeg} we compare the observed spectrum with the models showing the $H$-band region also used by \citet{smith_lambert_1986}.
The two $^{13}$CO lines visible in this figure show a divergent result. 
The 5-2 R48 line, used, among others, by Smith \& Lambert, is strongly affected by the neighboring OH line, but the depth indeed suggests a $^{12}$C/$^{13}$C ratio of more than 25.
On the other hand, the 5-2 R22 line strongly favors a lower ratio.
The plot clearly shows the weakness of the second-overtone lines and the according difficulty in determining the correct value.  The problem appears
to be the marginal usefulness of the observed spectrum.  A high-S/N spectrum in the $H$-band would settle this discrepancy. 

The literature value for $^{12}$C/$^{13}$C in V4028 Sgr \citep{tsuji_2008} is 48. 
As was the case for HR Peg and BD Cam the same FTS spectrum has been used in both studies.
However, the methods are different, with Tsuji's analysis being based on one single line. 
Tsuji notes that the uncertainty can be much larger in this case than the formal error, which is reproduced in Table  \ref{t:lit_values}.
A comparison between observed and a synthetic spectrum with $T_{\rm eff}$=3300\,K and log\,g=0.0 supports our $^{12}$C/$^{13}$C $\sim$ 15.
In the course of the modeling, the oxygen isotopic ratios of V4028 Sgr were confirmed within the error bars, with the synthetic spectra comparison favoring a slightly higher $^{16}$O/$^{17}$O value.

TV Aur has a literature value of $^{12}$C/$^{13}$C = 5$\pm$2 reported by \citet{smith_lambert_1990}.  The star was observed twice with 
the FTS, once in 1977 and again in 1987 (Table \ref{t:obslog}).  We have analyzed both spectra and assume the \citet{smith_lambert_1990} result is based 
on the second.  Both spectra have strong $^{12}$CO lines with the second-overtone lines reaching depths of 75\%.  By contrast the $^{13}$CO second-overtone 
lines are only marginally detectable in the spectra.  This requires that the $^{12}$C/$^{13}$C ratio be large and clearly excludes values $\lesssim$10.  The extraordinary strength of the $^{12}$CO second-overtone lines indicates the lines are saturated.  The available model atmospheres and spectrum synthesis techniques, such as those employed by \citet{smith_lambert_1990}, are not adequate for this star.    The low S/N in the $H$-band of both archival FTS spectra limits the accuracy of our results.  We confirmed that both FTS spectra were of TV Aur by checking the radial velocity against the Gaia value.  The isotopic ratios we derive from both spectra are similar and we adapt the average $^{12}$C/$^{13}$C = 40$\pm$10.

\subsubsection{Extreme $^{\it 16}$O/$^{\it 17}$O Values}

For the $^{16}$O/$^{17}$O (Figure \ref{fig:literatureOO}) we find good agreement with previous studies confirming the outcome of our analysis. In the case of HR Peg we note that the oxygen isotope values derived from a higher-S/N, higher-resolution 
spectrum agree with the literature values, while values derived from an inferior spectrum do not.
The very large values found for some stars have to be taken with caution since the C$^{17}$O lines become very weak in these cases and are even at an S/N=100 at the limit of detection.  For these very small values, small errors in placing the continuum make large 
differences in the isotopic ratio.  
Accordingly, we decided to show only lower limits in Table  \ref{t:meas_values} in these cases.
We do not present a plot comparing our result with literature values for the $^{16}$O/$^{18}$O because of the small number of literature values available.

\subsection{Comparison with synthetic spectra} \label{sect:synth}

Due to their large variability amplitudes and the dynamic nature of their atmospheres, the miras discussed in \citet{hinkle_et_al_2016} are not suitable for modeling with hydrostatic spectra \citep{lebzelter_et_al_2015b}.
However, the sample used in this paper consists of stars with comparably mild pulsations.
Therefore, we also derive isotopic ratios for oxygen by comparing our FTS spectra with synthetic spectra with a goal of comparing the methods.

We used hydrostatic atmospheres from the COMARCS grid of models by \citet{aringer_et_al_2016}, and computed spectra with the help of the COMA code \citep[][and references therein]{aringer_et_al_2016}. 
The effective temperatures were taken from the relation between $T_{\rm exc}$ and $T_{\rm eff}$ derived in Section \ref{sect:efftemp}.  The log g values were computed from Gaia data as discussed above.
From our sample of SRbs we selected a subset of objects (Table \ref{tab:comp_models}) with small uncertainties in the Gaia distances covering a wide range in effective temperature. 
Preference was given to stars whose parameters were close to the models provided by the COMARCS grid.
For all stars we assumed solar metallicity. Since our sample consists exclusively of bright, nearby stars, this assumption seems acceptable. Small deviations from solar metallicity do not
affect the outcome of the modelling process significantly.
For each of these sets of stellar parameters we computed synthetic spectra for various isotopic ratios of oxygen. 
The ratios of C/O and $^{12}$C/$^{13}$C were set to constant values of 0.3 and 10, respectively.
We did some tests to confirm that changing from $^{12}$C/$^{13}$C = 10 to 25 does not have a visible effect on the spectral regions investigated for the oxygen isotopes, nor do small changes in C/O.  The observational data were the telluric corrected spectra.  The continua were the same as for the curve-of-growth measurements.

\begin{table}
    \centering
    \caption{Comparison of the Results between Curve-of-growth Analysis and Spectrum Synthesis Analysis for the Ratios the Oxygen Isotopes.}
    \label{tab:comp_models}
    \begin{tabular}{ll|rrrr|rrrr}
    \hline
    \hline
         HD & Var   & \multicolumn{4}{c|}{CoG} & \multicolumn{4}{c}{spectrum synthesis}\\
         & Name & T$_{\rm eff}$ [K] & log\,g & $^{16}$O/$^{17}$O & $^{16}$O/$^{18}$O & T$_{\rm eff}$ [K] & log\,g & $^{16}$O/$^{17}$O & $^{16}$O/$^{18}$O\\
         \hline
         HD 18191 & RZ Ari & 3236 & 0.36 & 900$^{+850}_{-400}$ & 2200$^{+600}_{-400}$ & 3200 & 0.5 & 1600$\pm$300 & 5000$\pm$1000\\
         HD 39225 &  & 3709 & 1.04 & 600$^{+1100}_{-400}$ & 1000$^{+900}_{-500}$ & 3700 & 1.0 & 650$\pm$400 & 1300$\pm$400\\
         HD 71250 & BP Cnc & 3491 & 0.09 & 3300$^{+2700}_{-1900}$ & 2050$^{+1700}_{-900}$ & 3500 & 0.0 & 3500$\pm$1250 & 2500:\\
         HD 94705 & VY Leo & 3279 & 0.64 & 2200$^{+1400}_{-900}$ & 2400$^{+2200}_{-1100}$ & 3300 & 0.5 & 1900$\pm$300 & 2500$\pm$500\\
         HD 100029 & $\lambda$ Dra & 3758 & 1.04 & 1300$^{+1000}_{-600}$ & 650$^{+1000}_{-400}$ & 3800 & 1.0 & 1700$\pm$400 & 650$\pm$70\\
         HD 126327 & RX Boo & 2874 & -0.30 & 300$^{+300}_{150}$ & 250$^{+150}_{-100}$ & 2900 & -0.5 & 1000$\pm$200 & 1450$\pm$500\\
         HD 132813 & RR UMi & 3406 & 0.53 & $>$5000 & 550$^{+300}_{-200}$ & 3400 & 0.5 & 1700$\pm$450 & 725$\pm$100\\
         HD 133216 & $\sigma$ Lib & 3555 & 0.10 & $>$2000 & $>$2000 & 3600 & 0.0 & 3700$\pm$1400 & 1600$\pm$600\\
         HD 183439 & $\alpha$ Vul & 3768 & 1.06 & 2100$^{+1800}_{-1000}$ & 1900$^{+4900}_{-1300}$ & 3800 & 1.0 & 3700$\pm$2300 & 1900$\pm$200\\
         HD 184313 & V450 Aql & 3270 & 0.05 & 1000$^{+1700}_{-650}$ & 300$^{+200}_{-100}$ & 3300 & 0.0 & 2000$\pm$750 & 700$\pm$400\\
         \hline
    \end{tabular}
    \tablecomments{For the spectrum synthesis method, columns $T_{\rm eff}$ and log g give the parameters of the hydrostatic model used.}
\end{table}

\begin{figure}
    \centering
    \includegraphics[width=\columnwidth]{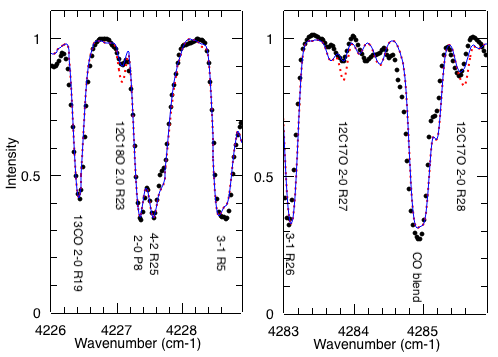}
    \caption{Model fit to the FTS spectrum of HR 4267. Big dots: observed spectrum. Left panel: part of the spectrum around the $^{12}$C$^{18}$O 2-0 R23 line; red dotted line: $^{16}$O/$^{18}$O=1000; blue solid line: $^{16}$O/$^{18}$O=2400. Right panel: $^{12}$C$^{17}$O 2-0 R27 and R28 lines; red dotted line: $^{16}$O/$^{17}$O=1000; blue solid line: $^{16}$O/$^{17}$O=2000.}
    \label{fig:HR4267_lines}
\end{figure}

An example of a model fit is shown in Figure \ref{fig:HR4267_lines} for the star HR 4267.
Table \ref{tab:comp_models} presents the resulting $^{16}$O/$^{17}$O and $^{16}$O/$^{18}$O ratios derived from spectral synthesis and compares them with the results from curve-of-growth analysis.
The spectrum synthesis approach focused on fitting the $^{16}$O/$^{17}$O 2-0 R25 to R32 lines and the $^{16}$O/$^{18}$O 2-0 R23 and R29 lines, respectively.
For each line we determined the most appropriate model ratios, and finally averaged over all lines to get the isotopic ratios and uncertainties listed in Table \ref{tab:comp_models}.
In cases in which one of these lines was obviously blended with a different stellar line or affected by incomplete telluric correction, it was rejected.

We find that for most stars of our subsample the values for the isotopic ratios agree for the two methods within the error bars.
It is noteworthy that the error bars of the two methods are quite different in size.
Here we have to consider that for the spectrum synthesis we did not include uncertainties in the stellar parameters. 
The error bars given for the spectrum synthesis method in Table \ref{tab:comp_models} reflect the line-to-line variation only since our comparison of the two approaches aimed at testing their consistency with each other. 

\begin{figure}
    \centering
    \includegraphics[width=\columnwidth]{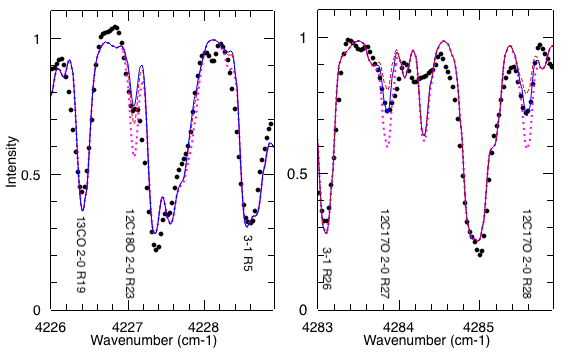}
    \caption{Model fit to the FTS spectrum of RX Boo. Big dots: observed spectrum. Left panel: part of the spectrum around the $^{12}$C$^{18}$O 2-0 R23 line; magenta dotted line: $^{16}$O/$^{18}$O=250; red dashed line: $^{16}$O/$^{18}$O=1000; blue solid line: $^{16}$O/$^{18}$O=2000. Right panel: $^{12}$C$^{17}$O 2-0 R27 and R28 lines; magenta dotted line: $^{16}$O/$^{17}$O=300; blue solid line: $^{16}$O/$^{17}$O=1000; red dashed line: $^{16}$O/$^{17}$O=2000.}
    \label{fig:rxboo_lines}
\end{figure}

Significant differences between the results from the two methods are visible for the stars HR 867, HR 5589, and RX Boo. 
The first two stars still show values of the same order of magnitude, which is the relevant information for the analysis we attempt in this paper.
For RX Boo (Figure \ref{fig:rxboo_lines}), however, the two methods lead to different conclusions, in particular in the case of $^{16}$O/$^{18}$O.
RX Boo is the coolest and most variable star in our subsample, and this finding may indicate a problem occurring for similar objects.
The temperature of this star is already at the lower end of our $T_{\rm exc}-T_{\rm eff}$ relation, and it is not clear whether the relation derived is still valid in this temperature regime.  RX Boo has a visual light amplitude around 2.5 mag. 
We have pointed out in Paper I and in \citet{lebzelter_et_al_2015b} that with increasing variability the assumption of a hydrostatic atmosphere is no longer appropriate. 
For consistency and considering the uncertainties of stellar modeling we did not account for in our exemplary study, we use the values derived by the curve-of-growth analysis in the following.
Clearly the curve-of-growth analysis has its shortcomings as well; thus, we refrain here from deciding on which method provides the more reliable results.
For stars with temperatures above 3000\,K and small variability amplitudes, our comparison shows that both methods lead to very similar conclusions.  For the cooler M giants the spectrum synthesis technique is increasingly questionable.

For the star HR 5603 the curve-of-growth analysis allowed us to derive lower limits for the two oxygen isotopic ratios only. 
In this case we will, as an exception, use the values derived from spectrum synthesis in the following analysis.

\section{DISCUSSION}\label{sect:discuss}

\subsection{Mass Estimates for SRb Variables Based on Isotopic Ratios}\label{sect:massestimate}

\begin{figure}
    \centering
    \includegraphics[width=\columnwidth]{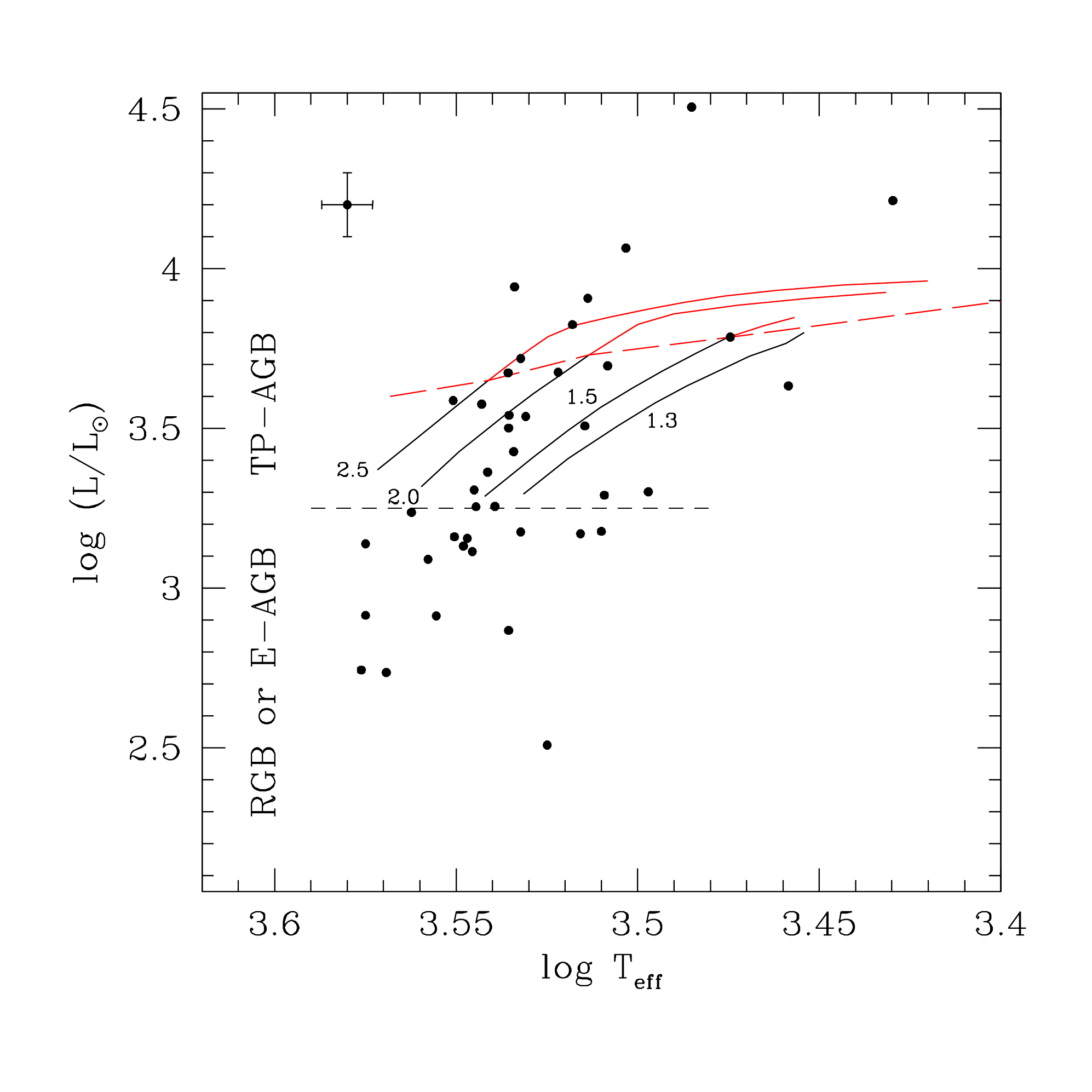}
    \caption{Hertzsprung-Russell-Diagram for our sample stars. Overplotted are evolutionary tracks from the FRUITY repository \citep{cristallo_et_al_2011} for ZAMS masses of 1.3, 1.5, 2.0 and 2.5 M$_{\sun}$, respectively. Portions of the tracks with recursive third dredge-up events are marked in red. The black dashed line shows the tip of the early-AGB phase. Long-dashed red line: onset of dredge-up
    episodes. Typical error bars are shown in the upper left corner.}
    \label{fig:HRD}
\end{figure}

\begin{figure}
    \centering
    \includegraphics[width=\columnwidth]{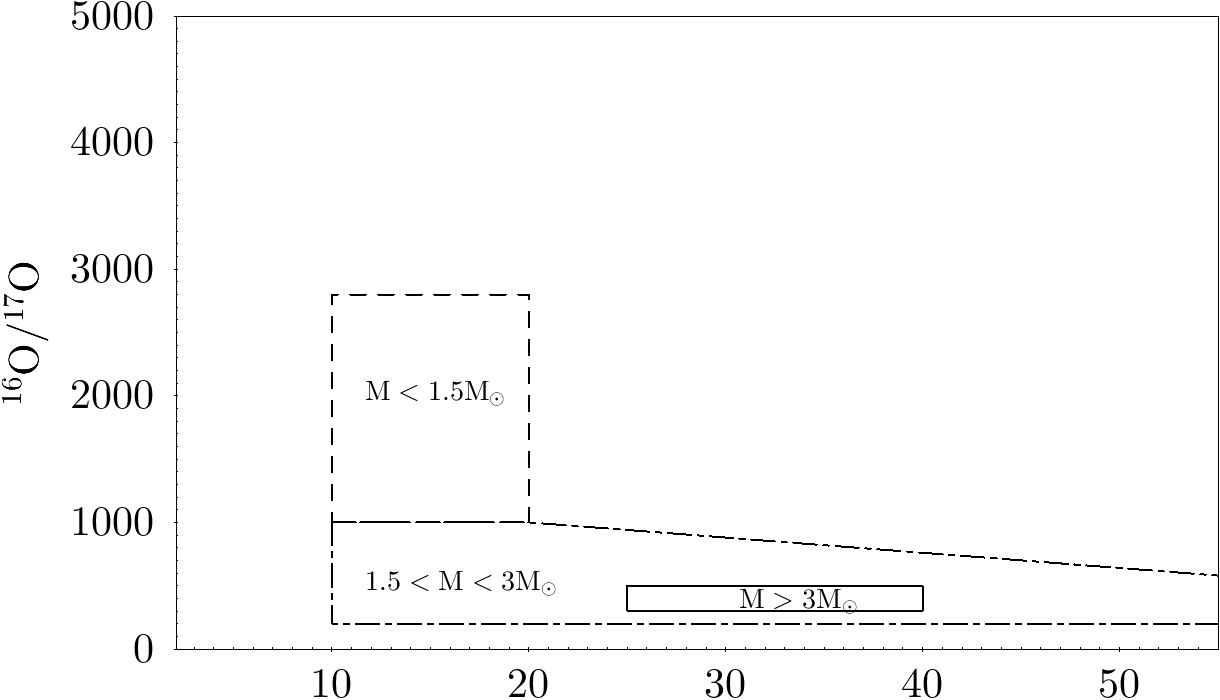}
    \includegraphics[width=\columnwidth]{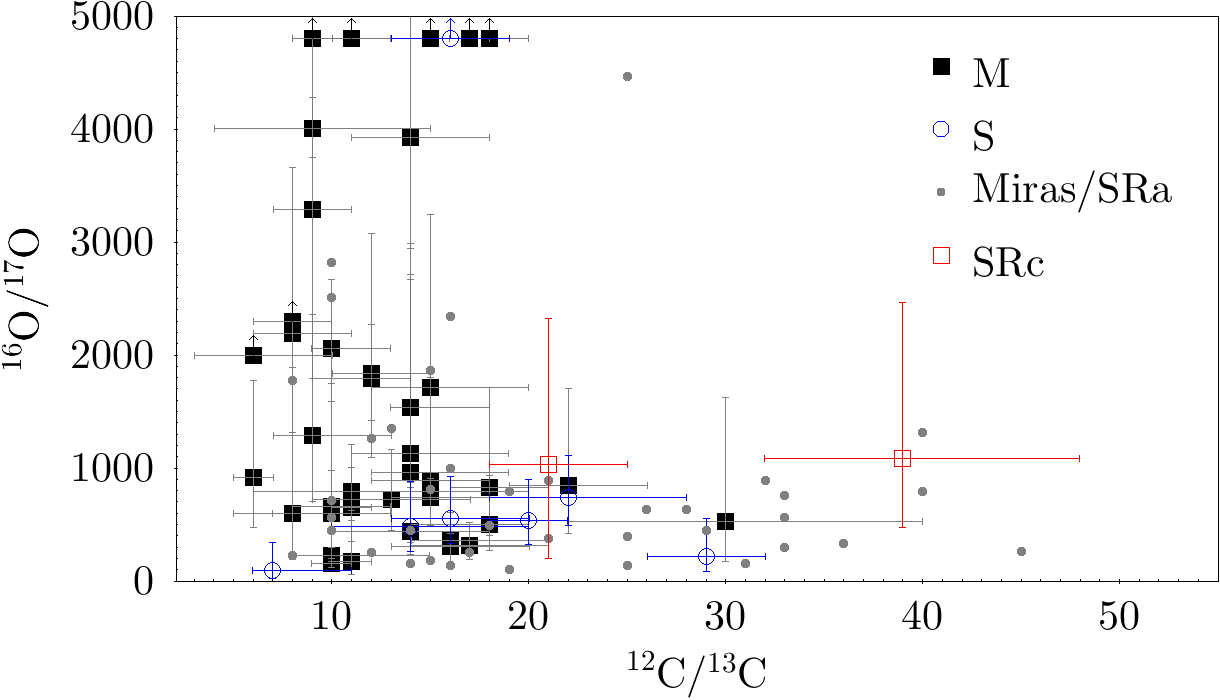}
    \caption{
    $^{16}$O/$^{17}$O vs. $^{12}$C/$^{13}$C for oxygen-rich LPVs. Filled boxes: SRbs of spectral type M (this paper); open circles: SRbs of type S (this paper); small filled circles: Miras and SRa variables (Paper I). Data points with arrows denote lower limits. The box marked by dashed lines indicates the location of stars below 1.5\,M$_{\sun}$ according to stellar evolution models, the area surrounded by dashed-dotted lines of stars between 1.5 and 3\,M$_{\sun}$. See text and Paper I for details.}
    \label{fig:C1213_O1617}
\end{figure}

Throughout the following discussion we assume that our sample consists of single stars.
In close binary systems, mass transfer could have changed the isotopic ratios on the surface significantly.
The carbon isotopic ratio is low for almost all stars of our sample. 
The values found are typical for a post-first-dredge-up composition, in many cases likely modified by extra mixing on the RGB to reach values below 10. 
This suggests that the SRbs studied here are, with a few exceptions, low-mass stars that have not experienced a third dredge-up either because they are still on the RGB/early AGB or because their envelope mass is too low for a third dredge-up to occur. 

To confirm the low masses of the sample stars an HRD was constructed (Figure \ref{fig:HRD}) using the values for $T_{\rm eff}$ and L from Section 4. 
Reddening was not taken into account.
A few stars in the sample lack reliable Gaia data and are not included.
Figure \ref{fig:HRD} also includes evolutionary tracks from the FRUITY database \citep{cristallo_et_al_2011} for ZAMS masses of 1.3, 1.5, 2.0 and 2.5 M$_{\sun}$, respectively.
Short time variations during a thermal pulse episode have been omitted for the sake of clarity, since stars are likely observed during the longer interpulse periods. 
The luminosity depends on the He core mass, which is very similar for the four tracks shown, while
the effective temperature is mainly defined by stellar mass and chemical composition (O- or C-rich). 
Obviously, our sample stars are located in the range spanned by the chosen evolutionary tracks except for a few bright objects.

The figure shows how the sample is divided between early-AGB/bright RGB stars and objects showing thermal pulses (dashed horizonal line).
Comparing the stars on the TP-AGB with the evolutionary tracks, we find that most of these objects have likely not experienced a third dredge-up event yet (objects below the red dashed line in Figure \ref{fig:HRD}).
Only six stars are clearly above the limit for third dredge-up: SW Vir, CY Cyg, TV Aur, HR Peg, V335 Hya, and V4028 Sgr (ordered by decreasing luminosity).
Half are S-type stars which will be discussed in detail in Section \ref{sect:s-stars}.

In Paper I a high $^{16}$O/$^{17}$O ratio was shown to be an indicator of a low main-sequence mass.
Models predict $^{16}$O/$^{17}$O=1000 stars with 1.5\,M$_{\sun}$, and $^{16}$O/$^{17}$O=2650 for 1\,M$_{\sun}$ (Paper I).
We show in Figure \ref{fig:C1213_O1617} the $^{16}$O/$^{17}$O ratio against the $^{12}$C/$^{13}$C ratio for our sample stars.
We conclude from that plot that the majority of our sample consists of stars of 1.5 to 2.5\,M$_{\sun}$, which have not yet experienced third dredge-up as indicated by their low isotopic ratio of carbon.
Since our sample was randomly selected among the small-amplitude LPVs, we propose that the above mentioned mass range is typical for SRbs/Lbs in the solar neighborhood.

The miras and SRa variables from Paper I 
have been included in  Figure \ref{fig:C1213_O1617}
with small symbols.
The bulk of the SRbs are located immediately to the left of the Miras and SRas.  Both groups 
of stars occupy the same range in $^{16}$O/$^{17}$O and hence cover the same range of mass.
Third dredge-up shifts the stars to the right in this diagram.
This suggests that SRbs are  progenitors of miras/SRas.  The SRbs have not yet reached the thermally pulsing AGB.

The two SRcs in our sample do not differ significantly from the rest of our sample.
Note that we could not compute luminosities for these two stars owing to the lack of a reliable Gaia parallax. 
Therefore, we cannot test their supergiant nature in this way. 
Their $^{12}$C/$^{13}$C ratios between 20 and 40 are comparable to what has been found for other red supergiants.
\citet{milam_et_al_2009} list a range of 8 to 46 for these luminous stars.
The $^{16}$O/$^{17}$O ratio is close to 1000 for both stars.
Studies on the oxygen isotopes in supergiants are rare. 
Values between 500 and 850 have been reported \citep{harris_lambert_1984a, geballe_et_al_1977}.

\begin{figure}
    \centering
    \includegraphics[width=\columnwidth]{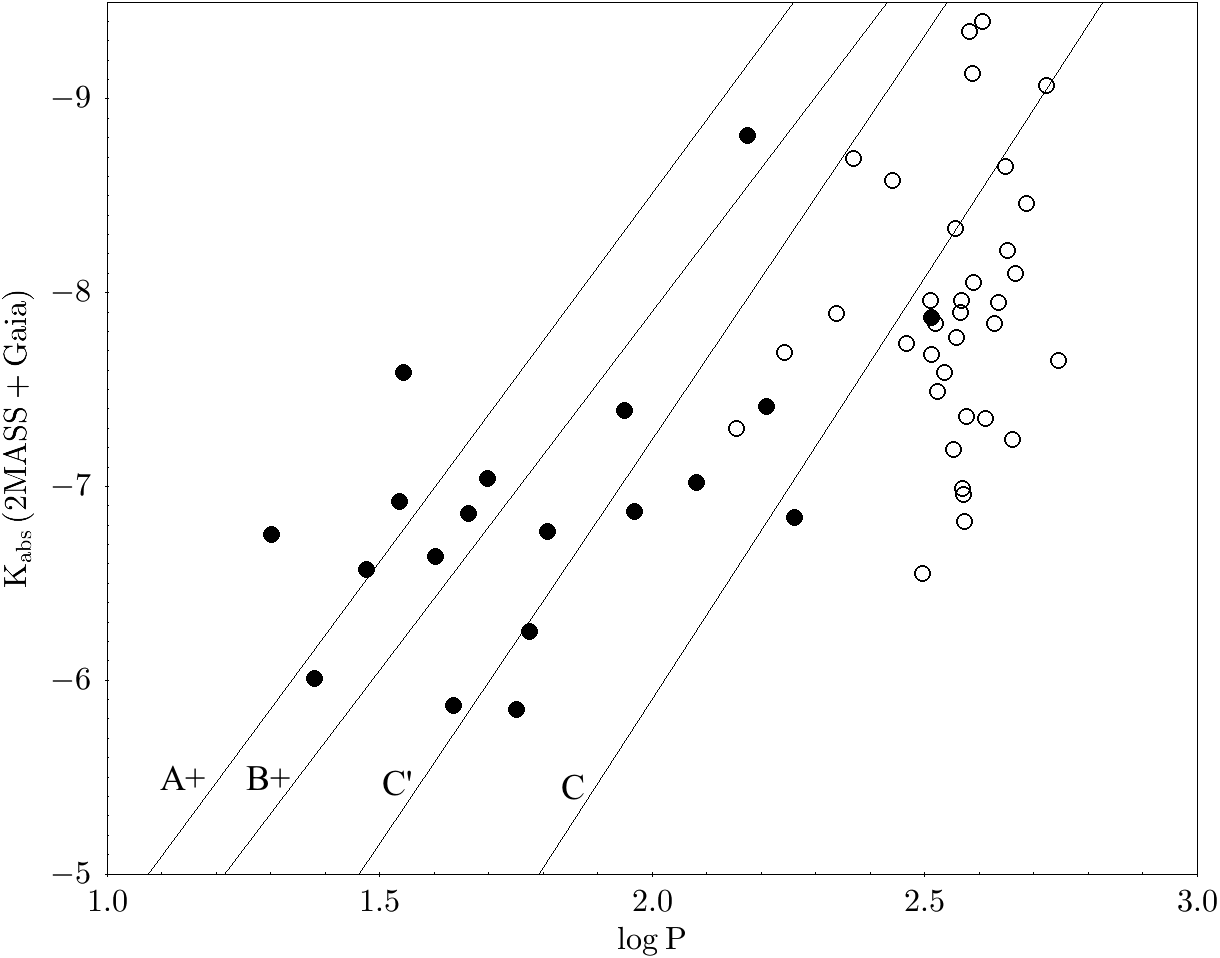}
    \caption{logP-K-diagram for the LPVs from Paper I (open circles) and this paper (filled symbols). Lines mark the logP-K-sequences derived from LMC data by \citet{ita_et_al_2004} corrected for the LMC distance \citep{degrijs_et_al_2017}. Sequences A+, B+, C', and C are shown from left to right.}
    \label{fig:pld}
\end{figure}

The evolutionary sequence suggested by the previous arguments is further supported by the location of the sample from this paper and the one from Paper I in a logP-K-diagram (Figure \ref{fig:pld}).
A large number of studies exist today revealing the presence of several parallel pulsation sequences in such a diagram. 
We indicated in Figure \ref{fig:pld} the ones derived by \citet{ita_et_al_2004} from LMC data.
AGB stars pulsate in the fundamental mode \citep[][sequence C]{wood_2015}  and overtones to the left the fundamental in this diagram.
\citet{trabucchi_et_al_2019} pulsation models for a 1.5 M$_\odot$ star have evolution tracks from the lower left to the upper right in this diagram, switching pulsation modes from overtone to fundamental.
This is consistent with Figure \ref{fig:pld} where the low-mass SRb stars evolve from lower left to upper right to become the low-mass miras of Paper I.

Our sample includes also a group of 11 stars with main-sequence masses around 1\,M$_{\sun}$ or less according to their $^{16}$O/$^{17}$O ratios.
Note that among these stars there are three with a very high $^{16}$O/$^{17}$O ratio. 
In such a case, the measured C$^{17}$O lines are already very weak, and while the order of magnitude of the ratio is likely correct, the exact value is very uncertain.
We show these points as limits.
Among these stars with very high $^{16}$O/$^{17}$O ratios is the S-type star DT Psc discussed in the next section.
There is only one mira with a comparably high $^{16}$O/$^{17}$O, namely U Ori (Paper I). 

\begin{figure}
    \centering
    \includegraphics[width=\columnwidth]{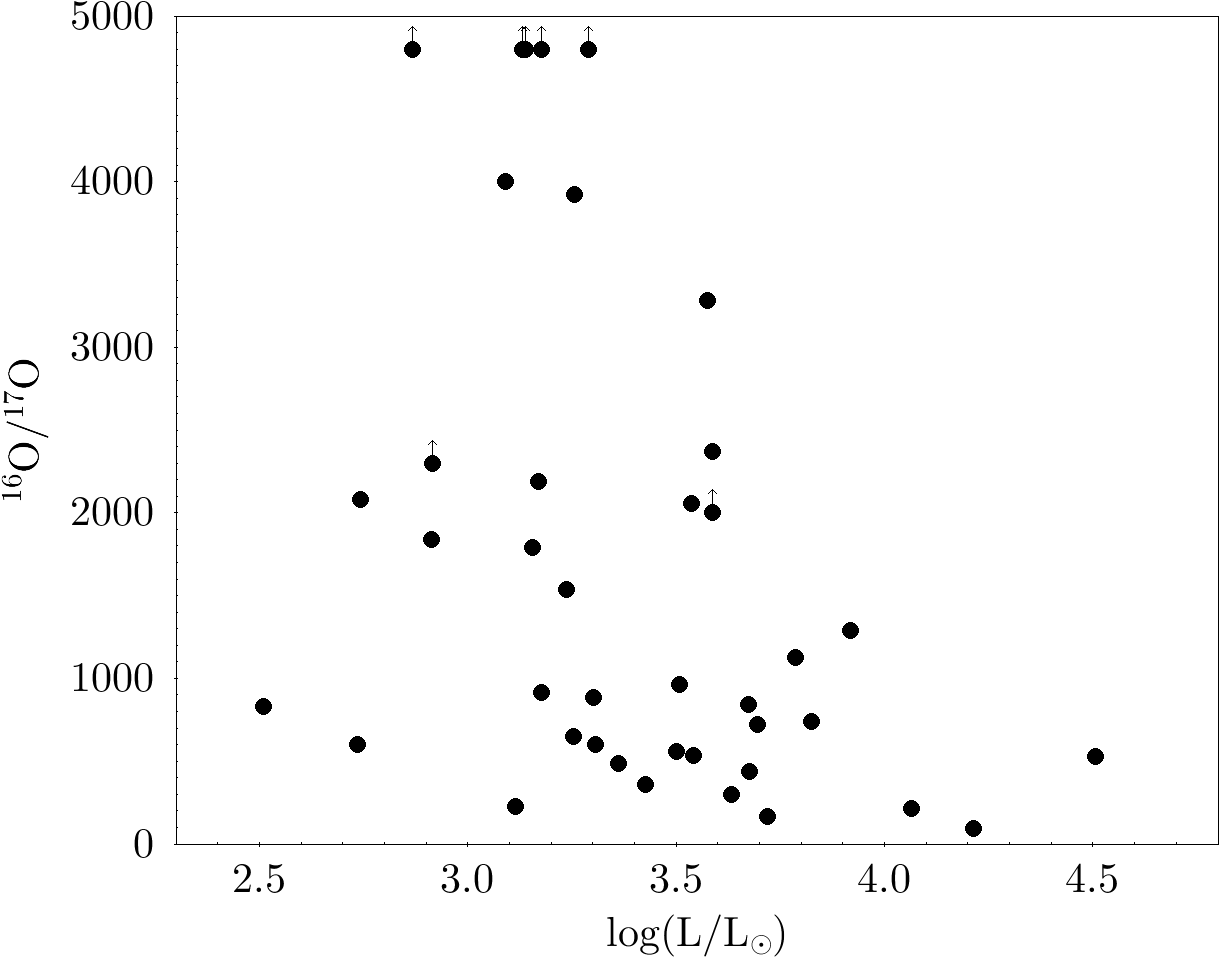}
    \caption{$^{16}$O/$^{17}$O ratio against luminosity. The sample shown here is limited to those objects for which a Gaia distance was available.  Data points with arrows denote lower limits.}
    \label{fig:logLO17}
\end{figure}

The $^{16}$O/$^{17}$O ratio is shown as a function of the luminosity in Figure \ref{fig:logLO17}.
No clear trend between these two quantities is visible. 
This complies with theoretical predictions, as, on the one hand, luminosity is dependent on He core mass, which is, as noted above, independent of the total mass in the case of low-mass stars.
As a consequence, luminosity is dependent on the evolutionary stage along the AGB.
On the other hand, the $^{16}$O/$^{17}$O ratio depends on the total stellar mass, but should stay constant during AGB evolution except for the case of extra-mixing being active during that phase.
At the highest luminosities, we only find low values of $^{16}$O/$^{17}$O, i.e. only high mass stars, as expected from the models.  

\begin{figure}
    \centering
    \includegraphics[width=\columnwidth]{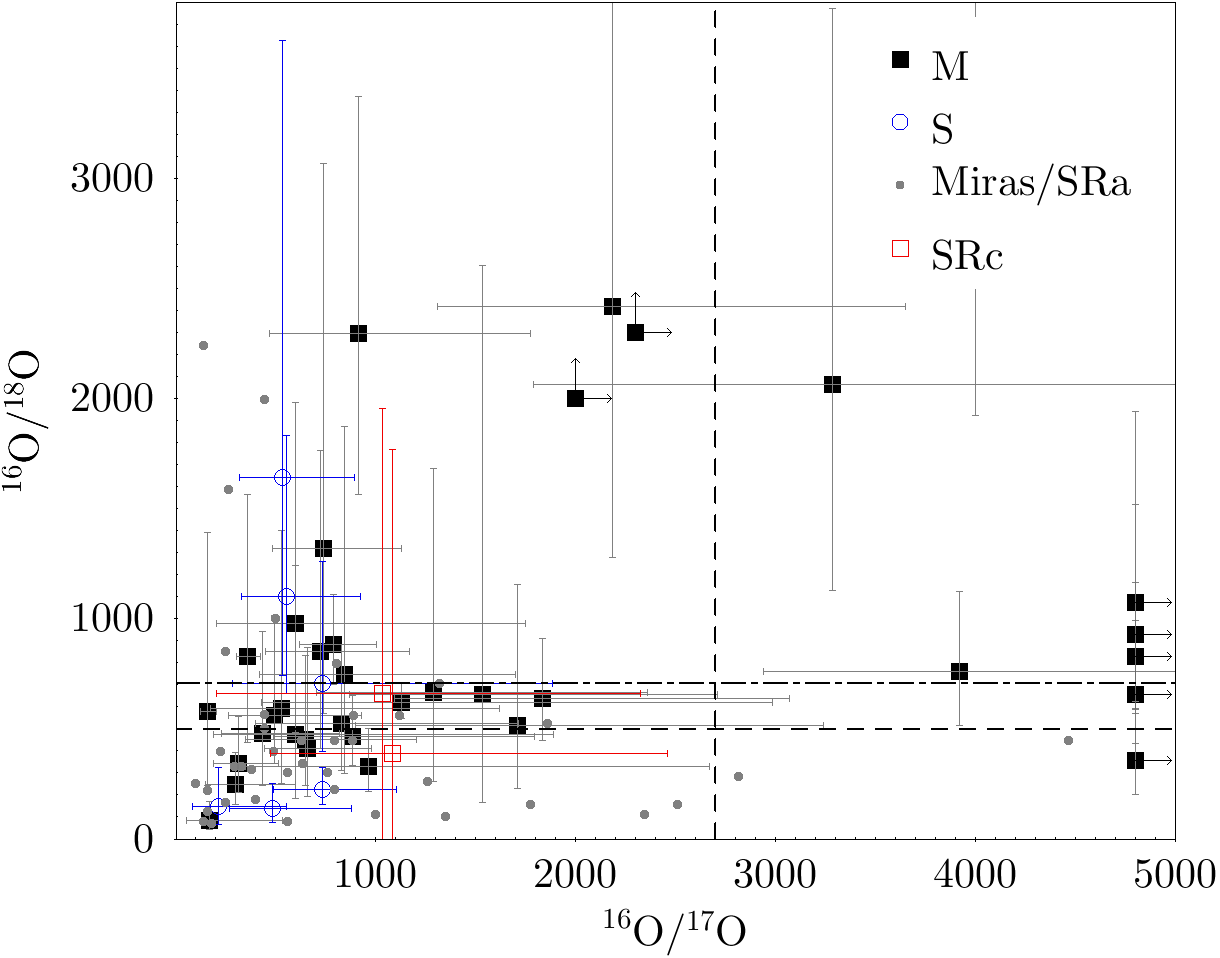}
    \caption{$^{16}$O/$^{17}$O vs. $^{16}$O/$^{18}$O for oxygen-rich LPVs. Same symbols as in Figure \ref{fig:C1213_O1617}.  Data points with arrows denote lower limits. The dashed lines indicate the solar values. The dashed-dotted line marks the $^{16}$O/$^{18}$O expected after first dredge-up.}
    \label{fig:O1617_O1618}
\end{figure}

In Figure \ref{fig:O1617_O1618} the $^{16}$O/$^{17}$O and $^{16}$O/$^{18}$O ratios are compared. 
Solar values are shown (dashed lines) as is the value for $^{16}$O/$^{18}$O expected after the first dredge-up \citep[dash-dot line, see][]{lebzelter_et_al_2015}.
Stars in the lower left corner, below the first dredge-up value and to the left of the solar $^{16}$O/$^{17}$O line,
comply with model expectations. 
They are more massive than the sun, therefore younger, and their lower $^{16}$O/$^{18}$O ratio suggests their formation from material enhanced in $^{18}$O owing to the galacto-chemical evolution \citep{prantzos_et_al_1996, kobayashi_et_al_2011, lebzelter_et_al_2015}.
A similar result was found for the bulk of the miras and SRas in Paper I.

Stars outside of this box require a different interpretation.
For those with $^{16}$O/$^{17}$O$>$3000 we may tentatively guess that they have masses of the order of 1\,M$_{\sun}$.
As far as Gaia parallaxes are available, we see from Figure \ref{fig:logLO17} that they have low luminosities and had probably not experienced a third dredge-up yet (Figure \ref{fig:HRD}).
If these stars are older than the sun, their primordial $^{16}$O/$^{17}$O ratio should have been higher following
galacto-chemical evolution models which predict a steep increase in the $^{16}$O/$^{17}$O ratio in the galaxy for ages older than the sun \citet[][Figure 4a]{prantzos_et_al_1996}.

If this scenario is correct, they should have large $^{16}$O/$^{18}$O values as well. 
We see in Figure \ref{fig:O1617_O1618} that two, maybe three stars comply with this condition.
The $^{16}$O/$^{18}$O values of the other stars with $^{16}$O/$^{17}$O$>$3000 are close to the first dredge-up value computed for a star of similar primordial composition as the sun.
These objects do not comply with predictions from stellar evolution models as either their $^{16}$O/$^{17}$O ratio is to high or their $^{16}$O/$^{18}$O ratio is too low. 
At this point we can only guess that their primordial composition deviated from the average ISM composition at the time of their formation, e.g. as the result of a nearby supernova.

One star in our SRb sample, RZ Ari, shows a high $^{16}$O/$^{18}$O combined with a comparably low $^{16}$O/$^{17}$O.
The latter suggests a mass higher than the sun.
For massive AGB stars, $\gtrsim$ 6 M$_\odot$, hot bottom burning decreases $^{12}$C/$^{13}$C and strongly increases $^{16}$O/$^{18}$O (see Paper I).
RZ Ari has a moderate luminosity, and we can thus exclude HBB or extra-mixing on the AGB affecting the surface composition.
We may speculate that the composition of RZ Ari results from mass transfer in a binary system.
The star had been part of a study of unresolved problem stars from the Hipparcos catalogue, i.e. stars with a suspected binarity, but Speckle interferometry did not produce conclusive results \citep{mason_et_al_1999}.
Among the miras, the star S Ori shows a very similar isotopic pattern.
Additional candidates both from Paper I and from the present sample have large error bars and are thus not included in this discussion.
We will come back to RZ Ari in \ref{sect:presolar}.

We conclude the discussion with the star SW Vir. 
This object has the highest luminosity in our sample and is clearly offset from the rest of the SRbs (Figure \ref{fig:HRD}). 
Taking its luminosity at face value, the star is either a supergiant or a massive AGB star (M$>$6 M$_{\sun}$).
Looking at the isotopic ratios, a red supergiant, that was a slow rotator on the main sequence, is the preferred hypothesis.
However, SW Vir shows a variability more typical of an AGB star than a supergiant. 
With its period of 150 d and $K_{abs}$, computed from the Gaia distance combined with the 2MASS $K$ magnitude, the star is located on overtone pulsation sequence A+ in Figure \ref{fig:pld}.
The light amplitude of more than 1 mag according to the GCVS is highly uncommon for a star on A+ \citep{kiss_bedding_2003}
raising doubts on the derived luminosity.
We note that the Gaia DR2 parallax is a factor of 2 smaller than the earlier Hipparcos measurement.
While we cannot exclude a supergiant classification from the existing data, the luminosity of SW Vir may indeed be considerably smaller.

\subsection{S-type stars} \label{sect:s-stars}

\begin{table*}
    \tablecolumns{7}
    \caption{Summary of the Properties and Isotopic Ratios of the S-stars Investigated. \label{tab:s-stars}}
    \begin{tabular}{llcccrrrrr}
       \hline
       \hline
       {HD } &  {Variable}  & {Spectral} & {Tc?} &  {Li?} & {$^{12}$C/$^{13}$C}  & {$^{16}$O/$^{17}$O} & {$^{16}$O/$^{18}$O} & {$^{12}$C/$^{13}$C} & {$^{16}$O/$^{17}$O}\\
               &  {star name} & {type}     &    {} &    {}  & \multicolumn{3}{c}{this paper/Paper I}   & \multicolumn{2}{c}{Literature\tablenotemark{a}} \\
       \hline
       7351 & DT Psc & S3+/2- & no\tablenotemark{b} & no & 16$^{+3}_{-3}$ & $>$8000 & \dots & 12, 24 & 2500, 3000\\
       22649 & BD Cam & S3.5/2 & no\tablenotemark{c} & no & 14$^{+6}_{-4}$ & 486$^{+394}_{-217}$ & 140$^{+115}_{-63}$ & 25, 34, 55 & 350, 2250\\
       30959 & o$^{1}$ Ori & M3SIII & yes\tablenotemark{d} & no & 16$^{+5}_{-2}$ & 360$^{+64}_{-54}$ & 827$^{+734}_{-389}$ & 10, 18, 25 & 480, 925\\
       49368 & V613 Mon & S3/2 & no\tablenotemark{e} & no\tablenotemark{k} & 16$^{+4}_{-3}$ & 556$^{+372}_{-223}$ & 1103$^{+732}_{-440}$ & 14 & 430 \\
       58521 & Y Lyn & M6S & yes\tablenotemark{c} & no & 39$^{+9}_{-7}$ & 1087$^{+1378}_{-608}$ & 387$^{+256}_{-154}$ & 27, 43 & 560\\
       64332 & NQ Pup & S4.5/2 & yes\tablenotemark{f} & \dots & 20$^{+2}_{-2}$ & 534$^{+359}_{-215}$ & 1643$^{+1985}_{-899}$ & 18 & 400\\
       163990 & OP Her & M6S & yes\tablenotemark{d} & no & 7$^{+4}_{-3}$ & 435$^{+386}_{-205}$ & 478$^{+461}_{-234}$ & 11, 20, 21 & 329, 360, 850\\
       198164 & CY Cyg & SC2/7.5 & yes\tablenotemark{b} & yes & 7$^{+4}_{-1}$ & 94$^{+248}_{-69}$ & \dots & 5, 5.6 & 350, 620\\
       216672 & HR Peg & S4+/1+ & yes\tablenotemark{c} & yes & 22$^{+6}_{-4}$ & 737$^{+368}_{-245}$ & 223$^{+102}_{-70}$ & 63, 71 & 2400\\
       218634 & GZ Peg & M4S & no\tablenotemark{g} & no & 11$^{+1}_{-1}$ & 169$^{+363}_{-115}$ & 86$^{+85}_{-43}$ & 30 & \dots\\
       260297 & DY Gem & S8,5 & no\tablenotemark{e} & \dots & 56$^{+23}_{-16}$ & 733$^{+1152}_{-448}$ & 704$^{+553}_{-309}$ & 30 & 570\\
              & TV Aur & S5/6 & yes\tablenotemark{e} & yes & 29$^{+3}_{-3}$ & 213$^{+339}_{-131}$ & 149$^{+176}_{-81}$ & 5 & 240\\
       \hline
       1967 & R And & S5-7/4-5e & yes\tablenotemark{h} & no & 40$\pm$15 & 1320$\pm$300 & 710$\pm$150 & 24, 33 & \dots \\
       29147 & T Cam & S5-6/5e & yes\tablenotemark{b} & yes & 31$\pm$10 & 160$\pm$70 & 220$\pm$90 & \dots & \dots \\
       185456 & R Cyg & S4-8/6e & yes\tablenotemark{b} & yes & 29$\pm$10 & 450$\pm$120 & 500$\pm$200 & 24, 26, 34 & 860\\
       187796 & $\chi$ Cyg & S6-9/1-2e & yes\tablenotemark{b} & no & 36$\pm$4 & 330$\pm$100 & 330$\pm$70 & 39, 40 & \dots\\
       209890 & RZ Peg & SC5-9/9-e & yes\tablenotemark{b,j} & yes & 9$\pm$5 & \dots & \dots & 12 & \dots\\ 
       \hline
    \end{tabular}
    (a) {References see Table  \ref{t:lit_values} and Table  5 of Paper I.} (b) {\citet{little_et_al_1987}.} (c) {\citet{smith_lambert_1988}.} (d) {\citet{lebzelter_hron_1999}.} (e) {\citet{smith_lambert_1990}.} (f) {\cite{shetye_et_al_2018}.} (g) {\citet{vaneck_et_al_1998}.} (h) {\citet{jorissen_et_al_1993}.} (j) {\citet{vanture_et_al_2007} give log\,$N$(Tc)$\leq$0.4.}
    (k) {No Li line visible in ELODIE library spectrum \citep{prugniel_et_al_2007}.
    The final five objects are miras/SRas from Paper I. Information on the presence of a Li line is from \citet{vanture_et_al_2007}.}
\end{table*}

Stars of spectral type S show a strong enhancement of $s$-process elements on their surface. There are two processes known to form S-type stars, either by third dredge-up on the AGB (intrinsic S-stars) or by mass transfer from a further evolved companion in a binary system \citep[extrinsic S-stars, see for additional discussion][]{vanture_et_al_2007}. 
The intrinsic classification depends on detection of the radioactive s-process element Tc. Since the half-life time of its longest-living isotope is only 2$\times$10$^{6}$\,yr, any Tc detectable on the surface must have been produced and dredged to the surface recently. 
In contrast, binarity is the criterion to identify an extrinsic S-star \citep{brown_et_al_1990}. 

Intrinsic S-stars must have already experienced at least one third dredge-up event. This should be detectable in the $^{12}$C/$^{13}$C ratio. A reasonable expectation is that the intrinsic S-stars will have $^{12}$C/$^{13}$C intermediate between values for tip AGB oxygen-rich stars (8$\lesssim$$^{12}$C/$^{13}$C$\lesssim$45, Paper I) and carbon stars
\citep[30$\lesssim$$^{12}$C/$^{13}$C$\lesssim$70,][]{lambert_et_al_1986}.  The data on the S and MS stars in our sample and in Paper I are summarized in Table  \ref{tab:s-stars}. 
Drawing on the work of \citet{vanture_et_al_2007} we have added information on the presence of Tc (column 4) and on Li (column 5).
Lithium is destroyed by H-burning on the main sequence and should not be present after the first dredge-up. However, observations clearly indicate its presence all along the RGB \citep{charbonnel_et_al_1999, lebzelter_et_al_2012}. 
Extra-mixing on the RGB can be invoked to explain the Li.  Li also can be produced in stars with masses above 3\,M$_{\sun}$ by HBB via the Cameron-Fowler-mechanism. 
Examples for the latter process have been found in Li-rich, luminous S-stars in the Magellanic Clouds \citep{smith_lambert_1989}.

Our sample includes 15 S and MS stars and two SC stars. 12 of them show Tc lines and are thus intrinsic S-stars. Among them is o$^{1}$ Ori, a star that is also known to be a binary \citep[e.g.][]{lebzelter_hron_1999}. The average $^{12}$C/$^{13}$C ratio of this group is 20, consistent with a mild enhancement of $^{13}$C by third dredge-up. 
However, the scatter among the intrinsic S-stars  is quite large. 
The two SC stars in this group show a much lower value of 7, which indicates a post first dredge-up composition. \citet{abia_et_al_2017} have pointed out that SC stars show a very peculiar abundance pattern in general, possibly close to barium stars \citep{dominy_et_al_1986}, which are binaries. In this scenario, the presence of Tc is difficult to understand.
We note that \citet{vanture_et_al_2007}, using measurements by \citet{abia_wallerstein_1998}, give only upper limits for Tc in our two SC stars, CY Cyg and RZ Peg, casting doubt on their classification as intrinsic S-stars.
However, \citet{jorissen_et_al_1998} did radial velocity monitoring for the two stars and could not find convincing evidence for orbital motion.

Removing the SC stars from the intrinsic S-stars leaves only OP Her and NQ Pup with a low carbon isotopic ratio in this group. In both cases, previous papers give similar or somewhat higher numbers.  Clearly OP Her and NQ Pup show only a mild enhancement of $^{13}$C, if at all. Alternatively, extra-mixing may have occurred during the AGB evolution of these stars \citep{lebzelter_et_al_2008}, which could have reduced their carbon isotopic ratio. 

The five S-stars without Tc in our sample have $^{12}$C/$^{13}$C ratios between 7 and 23. Results from the literature for these stars give values between 12 and 55. 
This suggests that this group consists of stars before and after the onset of third dredge-ups. In the latter ones the number of dredge-ups or the mixing efficiency may have been too low to bring a detectable amount of Tc to the surface. 

Besides Tc, the element Nb can be used to discriminate between intrinsic and extrinsic S-stars \citep[e.g.][]{neyskens_et_al_2015}. Extrinsic S-stars should
show an enhancement of Nb, while intrinsic S-stars should
be depleted. For five stars in our sample we found measurements of Nb abundance in the literature, namely, BD Cam, V613 Mon, DT Psc, HR Peg, and NQ Pup \citep{neyskens_et_al_2015, karinkuzhi_et_al_2018}. In all five cases, the classification derived from the presence or absence of Tc is confirmed.

There is no correlation between $^{12}$C/$^{13}$C and the presence of Li. The occurrence of Li and Tc in one star may be an indicator of an intermediate-mass star with proper conditions for HBB in its interior. 
There are two candidates in our sample, TV Aur and T Cam, which not only show both elements but also have a low $^{16}$O/$^{17}$O ratio. These stars would then be higher-mass AGB stars. This is supported by the high luminosity for TV Aur (Table \ref{t:texc}).  In the event of hot bottom burning $^{16}$O/$^{18}$O is expected to be large, but this is not so for either star.

Except for DT Psc, the $^{16}$O/$^{17}$O ratios of the S-type stars are around or below 1000, suggesting masses above 1.5\,M$_{\sun}$.
There is no systematic difference between intrinsic and extrinsic S-stars.
To explore the relation between carbon and oxygen isotopes and the two classes of S-stars conclusively, a study on a larger sample of objects is necessary.

\subsection{The role of LPVs in the cosmic matter cycle} \label{sect:presolar}

\begin{figure}
    \centering
    \includegraphics[width=\columnwidth]{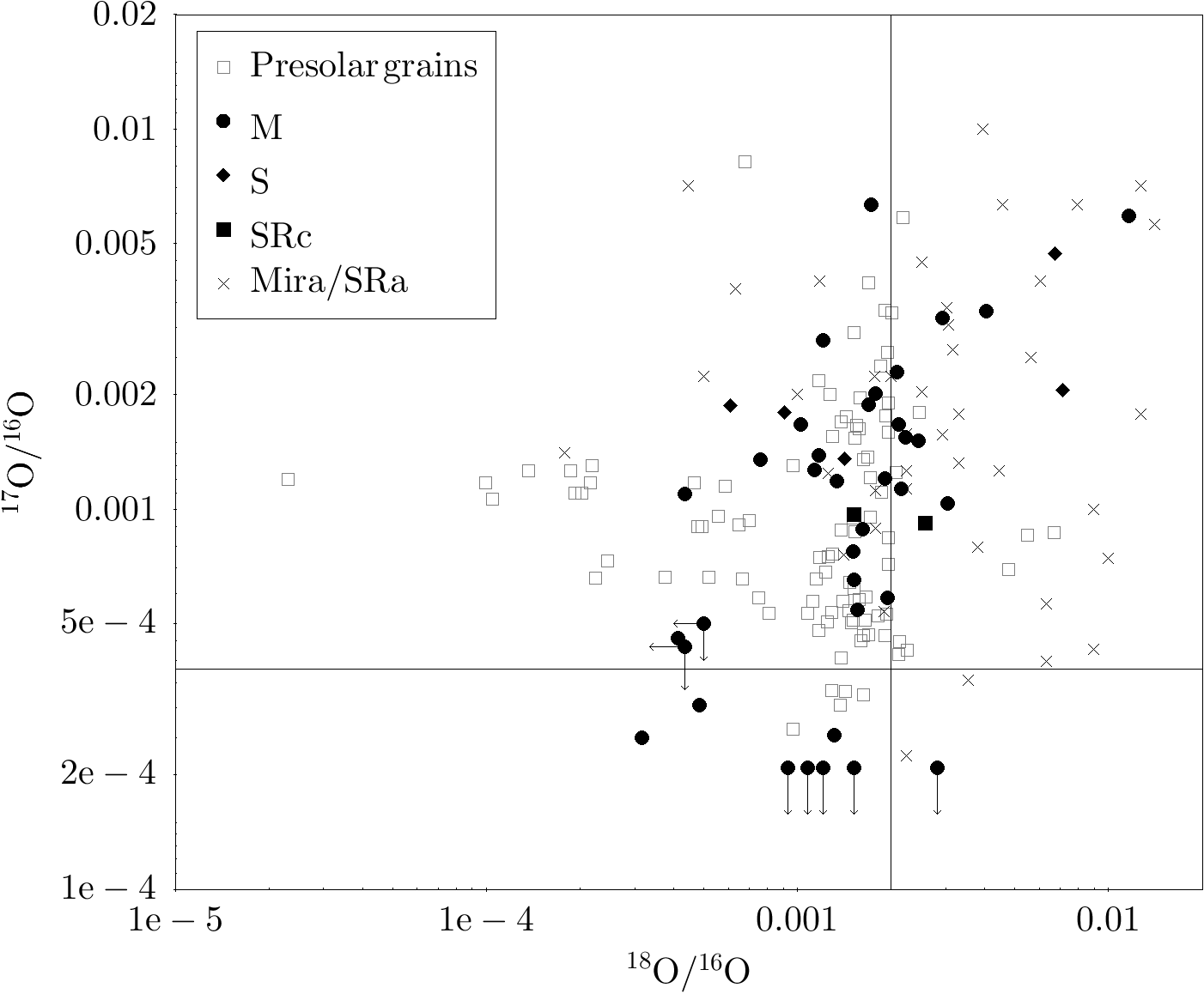}
    \caption{$^{17}$O/$^{16}$O vs. $^{18}$O/$^{16}$O comparing results from AGB stars with presolar grains (open boxes). Filled circles and diamonds mark M- and S-type variables from this paper, respectively. Filled boxes refer to the two SRcs in our sample. Crosses mark miras and SRas from Paper I. The two lines mark the isotopic ratios for the sun. Isotopic ratios are plotted reverse compared to the rest of the paper.  Symbols with arrows mark upper limits.}
    \label{fig:presolar}
\end{figure}

Presolar grains, i.e. dust grains that survived the formation process of the solar system without modification and are found in pristine meteorites, are indicators of the various contributors to the cloud from which the sun formed \citep{zinner_1998}.
The majority of the oxygen-rich grains are believed to originate in AGB stars because of the isotopic signature they show.
In Paper I we compared the oxygen isotopic ratios of miras and SRas with collected data for presolar grains from \citet{nittler_et_al_2008}.
In this comparison it is important to remember that the contributors of dust to the presolar nebula came from stars that had an AGB phase at least 5 billion years ago.

In Figure \ref{fig:presolar} we plotted the two oxygen isotopic ratios for a sample of presolar grains, the sample from Paper I, and our current sample.
The isotopic ratios have been inverted in this discussion to comply with the standard notation used for presolar grains.
The bulk of the miras and SRas have larger than solar $^{18}$O/$^{16}$O values (Paper I). 
This was expected as most of these AGB stars must have formed later than the contributors of the presolar grains with their $^{18}$O/$^{16}$O ratio modified by the galactic evolution.
The majority of the stars presented in this paper, SRbs and Lbs, are found much closer to the solar value and occupy almost the same area in the diagram as the so-called group 1 presolar grains.
This difference between the Paper I stars and the SRbs can be explained by the depletion of $^{18}$O during stellar evolution.
The $^{12}$C/$^{13}$C ratio we find in the SRbs is typically around 10 or lower (Figure \ref{fig:C1213_O1617}).
Such values are an evidence of RGB extra-mixing and implies temperatures of 23 to 24\,MK in the area reached by the first dredge-up. 
At such conditions, $^{18}$O is expected to be depleted as it is seen in our observations.
In $^{17}$O/$^{16}$O, an identical range is covered by this group of stars as by the presolar grains.

As pointed out by \citet{nittler_2009}, stellar evolution models attribute group 1 stars to the dusty winds from O-rich AGB stars, and our observations nicely confirm this. 
When simply associating the abscissa in this diagram with an indicator for the age of a star, many members of our sample would have an age between the sources of the presolar grains and the sun.
Thus, they should have a mass somewhat above 1\,M$_{\sun}$ with the associated $^{17}$O/$^{16}$O ratios as seen in Figure \ref{fig:presolar} (compare Section \ref{sect:massestimate}).
In accordance with this picture, the SRbs in the upper right corner of Figure \ref{fig:presolar} show $^{18}$O/$^{16}$O ratios similar to the younger miras and SRas from Paper I. 
However, we have to be aware that the primordial amounts of the oxygen isotopes are not equally distributed within our Galaxy resulting in scatter due to the different places of origin of our sample stars.
\citet{wilson_rood_1994} showed that the interstellar $^{18}$O/$^{16}$O ratio rather continuously changes by a factor of two between the Galactic centre and the local ISM.
Furthermore, stars may have been enriched during their formation by material from massive stars \citep{wouterloot_et_al_2008}.

There are a few stars that do not follow the above-mentioned general trend:
We notice four stars with $^{18}$O/$^{16}$O close to solar but $^{17}$O/$^{16}$O subsolar, i.e. one would assume that these stars should have a lower mass than the sun, but they were formed at a similar time as the majority of our sample.
There is also one mira found in this part of the diagram. 
We mentioned this strange group of objects already in Section \ref{sect:massestimate} and speculated about a potential modification of their composition by a nearby supernova exploding before their birth.
The few grains visible close to the solar isotopic ratios, named group 3, are associated with an origin in the interstellar medium.
However, it may be that such grains stem from this potentially rare group of stars.

One star of our sample, RZ Ari, shows oxygen isotopic ratios similar to grains in group 2. 
That group can be explained by AGB stars experiencing extra-mixing on the AGB or hot bottom burning (both processes reduce their $^{18}$O/$^{16}$O, see Paper I).
The $^{17}$O/$^{16}$O ratio as well as the derived luminosity do not favor a mass $>$ 3\,M$_{\sun}$.
A strong Li line, found by \citet{merchant_1967} has its origin probably in processes on the RGB \citep[e.g.][]{lebzelter_et_al_2012}.
In Paper I we noted that the star SV Cas belongs to this group, too. These two targets are interesting for further studies on the phenomenon of extra-mixing on the AGB.
We may speculate that the four stars at similar $^{18}$O/$^{16}$O but lower $^{17}$O/$^{16}$O are also affected by extra-mixing.

\section{CONCLUSIONS}

In this paper we presented isotopic ratios of the two key elements C and O for a sample of mainly O-rich long-period variables with small amplitudes.
We demonstrated that these isotopic ratios can be measured accurately with the classical curve-of-growth method.
Comparison with other methods in the form of literature values and model spectra computed in the course of this study illustrates that the various methods lead to similar results except for very cool and highly variable objects.
Our comparison focused on O-rich stars only.  Due to the low excitation temperatures of the cool carbon stars, an extension of this technique to these stars is a complex undertaking and beyond the scope of the present work.

Combined with the results from Paper I, our study allows a broad view of the isotopic composition of the whole range of LPVs.
The three ratios $^{12}$C/$^{13}$C, $^{16}$O/$^{17}$O, and $^{16}$O/$^{18}$O are related to the evolutionary state, the mass and the primordial composition of these stars and thus allows insight into these key parameters.
We find that most small-amplitude LPVs have main-sequence masses between 1.3 and 2.5 M$_{\sun}$. 
An evolutionary scenario from these small-amplitude variables to the large amplitude miras, suggested also by their location in the HRD and in a period-luminosity diagram, is in very good agreement with the findings from the isotopic ratios of C and O.

The majority of the isotopic patterns observed in SRVs and Miras complies well with model predictions. 
They can be explained in terms of mass and evolutionary differences and signatures of galacto-chemical evolution, so that we can attribute a mass range, an age, and an evolutionary phase to the various groups among the LPVs.
Results are consistent with conclusions derived from comparison with stellar evolution tracks and pulsational properties.

Finally, we contrasted our results to the findings from presolar grains.
Grain data on isotopic ratios reflect the composition of dust in the pre-solar nebula and should provide an imprint of the dust-producing stars contributing to the origin of the solar system. 
The LPVs studied here and in Paper I agree well with presolar grains in $^{17}$O/$^{16}$O, suggesting that the majority of these grains stem from AGB stars in the 1.3 to 2.5 M$_{\sun}$ range.
While miras and SRas show an offset from presolar grains in $^{18}$O/$^{16}$O (as noted in Paper I), the SRbs and irregular variables investigated in this paper do not show such a shift.
We suggest that this is due to a mass difference between the miras and the SRbs sample and as a consequence a depletion of $^{18}$O.
A few data points in our comparison do not follow the expectations from standard models and will require further investigation.

\acknowledgments 
The authors are indebted to Bernhard Aringer for providing access to his model atmosphere code and for his support with computing the synthetic spectra used in this paper. Sharon Hunt provided assistance with the citations, Overleaf, and ORCID. 
NSF's National Optical-Infrared Astronomy Research Laboratory, which has superseded the National Optical Astronomy Observatory (NOAO), is operated by the Association of Universities for Research in Astronomy under cooperative agreement with the National Science Foundation.  KH thanks the NOAO Office of Science for supporting his research.  
This work made use of data from the European Space Agency (ESA) mission $Gaia$ (\url{https://www.cosmos.esa.int/gaia}), processed by the Gaia Data Processing and Analysis Consortium (DPAC).  
Funding for the DPAC has been provided by national institutions, in particular the institutions participating in the $Gaia$ Multilateral Agreement.
This research was facilitated by the SIMBAD database, operated by CDS in Strasbourg, France, and NASA's Astrophysics Data System Abstract Service.  
We acknowledge with thanks the variable star observations from the AAVSO International Database contributed by observers worldwide and used in this research.


\clearpage

\end{document}